  \RequirePackage{fix-cm}
  \documentclass[twocolumn]{svjour3}          
  \smartqed  
  \usepackage{graphicx}
  \usepackage{natbib}
  \usepackage{multirow}
  \usepackage{booktabs}
  \usepackage[title]{appendix}
  \usepackage{url}
  \usepackage{amssymb}
  \usepackage{stmaryrd}
  \usepackage{amsmath}
  \usepackage[justification=centering,caption=false]{subfig}
  \usepackage{hyperref}
  
  \usepackage[final]{changes}

  \newcommand{\ie}{i.e.}
  \newcommand{\eg}{e.g.}
  
  \long\def\comment#1{}
  
\begin{document}
  
  \title{Comparison of Full-Reference Image Quality Models for Optimization of Image Processing Systems}
  
  
  
  \author{Keyan Ding$^{1}$ \and
          Kede Ma$^{1}$ \and
          Shiqi Wang$^{1}$ \and
          Eero P. Simoncelli$^{2}$
  }
  
  
  \institute{Keyan Ding \at
                \email{keyan.ding@my.cityu.edu.hk}
             \and
             Kede Ma (Corresponding author)\at
                \email{kede.ma@cityu.edu.hk}
             \and
              Shiqi Wang  \at
                \email{shiqwang@cityu.edu.hk}   
              \and
              Eero P. Simoncelli  \at
                \email{eero.simoncelli@nyu.edu}    
               \\
               \\
              $^{1}$ Department of Computer Science, City University of Hong Kong, Kowloon, Hong Kong  \\
              $^{2}$ Howard Hughes Medical Institute, Center for Neural Science, and Courant Institute of Mathematical Sciences, New York University, New York, USA
             }
  
  \date{}

  \maketitle
  
  \begin{abstract}
  The performance of objective image quality assessment (IQA) models has been evaluated primarily by comparing model predictions to human quality judgments. Perceptual datasets \comment{(\eg, LIVE and TID2013)} gathered for this purpose have provided useful benchmarks for improving IQA methods, but their heavy use creates a risk of overfitting. Here, we perform a large-scale comparison of IQA models in terms of their use as objectives for the optimization of image processing algorithms. Specifically, we use eleven full-reference IQA models to train deep neural networks for four low-level vision tasks: denoising, deblurring, super-resolution, and compression.  Subjective testing on the optimized images allows us to rank the competing models in terms of their perceptual performance, elucidate their relative advantages and disadvantages in these tasks, and propose a set of desirable properties for incorporation into future IQA models.
  
   
  \keywords{Image quality assessment \and Perceptual optimization \and Performance evaluation}
  \end{abstract}

  \section{Introduction}\label{intro}
  The goal of objective image quality assessment (IQA) is the construction of computational models that predict the perceived quality of visual images. 
  IQA models are generally classified according to their reliance on the availability of an original reference image. Full-reference methods compare a distorted image to the complete reference image,  reduced-reference methods require only partial information about the reference image, and  no-reference (or blind) methods operate solely on the distorted image. The standard paradigm for testing IQA models is to compare them to human quality ratings of distorted images, which have been made available in datasets such as LIVE~\citep{LIVE} and TID2013~\citep{Ponomarenko201557}. However, excessive reuse of these test sets during IQA model development may lead to overfitting, and as a consequence,  poor generalization to images corrupted by distortions that are not present in the test sets (see Table~\ref{tab:srcc_cmp2}).
  
  A highly promising but relatively under-studied application of IQA measures is to use them as objectives for the design and optimization of new image processing algorithms. The parameters of image processing methods are usually adjusted to minimize the mean squared error (MSE), the simplest of all fidelity metrics, despite the fact that it has been widely criticised for its poor correlation with human perception of image quality~\citep{girod1993what}. Early attempts at perceptual optimization using the \textit{structural similarity} (SSIM) index \citep{wang2004image} in place of MSE achieved perceptual gains in applications of image restoration~\citep{channappayya2008ssim}, wireless video streaming~\citep{vukadinovic2009trade}, video coding~\citep{wang2011ssim}, and image synthesis~\citep{snell2017learning}. A recent publication used perceptual measures based on pre-trained deep neural networks (DNNs) for optimization of super-resolution results~\citep{johnson2016perceptual}, although these have not been tested against human judgments.
   
  In this paper, we systematically evaluate a large set of full-reference IQA models in the context of perceptual optimization. To determine their suitability for optimization, we first test the models on recovering a reference image from a given initialization by optimizing the model-reported distance to the reference. For many IQA methods, we find that the optimization does not converge to the reference image, and can generate severe distortions. These optima are either local, or global but non-unique. We select eleven optimization-suitable IQA models as perceptual objectives, and use them to optimize DNNs for four low-level vision tasks - image denoising, blind image deblurring, single image super-resolution, and lossy image compression.  Extensive human perceptual tests on the optimized images reveal the relative performance of the competing models. Moreover, inspection of their visual failures indicates limitations in model design, providing guidance for the development of future IQA models.


  \section{Taxonomy of Full-Reference IQA Models}\label{sec:bg}
  Full-reference IQA methods can be broadly classified into five categories:
  \begin{itemize}
      \item {\it Error visibility methods} apply a distance measure directly to pixels (e.g., MSE), or to transformed representations of the images. The MSE in particular possesses useful properties for optimization (e.g., differentiability and convexity), and when combined with linear-algebraic tools, analytical solutions can often be obtained. For example, the classical solution to the MSE-optimal denoising problem (assuming a translation-invariant Gaussian signal model) is the Wiener filter~\citep{wiener1950}. Given that MSE in the pixel domain is poorly correlated with perceived image quality, many IQA models operate by first mapping images to  more perceptually appropriate representations~\citep{safranek1989perceptually,daly1992visible,lubin1993use,andrewdctune1993,teo1994perceptual,watson1997visibility,larson:011006,laparra2016perceptual}, and  measuring MSE within that space.
      
      \item {\it Structural similarity (SSIM) methods} are constructed to measure the similarity of local image ``structures'', often using correlation measures. The prototype is the SSIM index~\citep{wang2004image}, which combines similarity measures of three conceptually independent components - luminance, contrast and structure.  It has become a {\em de facto} standard in the field of perceptual image processing, and provided a prototype for subsequent IQA models based on feature similarity~\citep{zhang2011fsim}, gradient similarity~\citep{liu2012image}, edge strength similarity~\citep{zhang2013edge}, and saliency similarity~\citep{zhang2014vsi}. 
      
      \item {\it Information-theoretic methods} measure some approximation of the mutual information between the perceived reference and distorted images. Statistical modeling of the image source, the distortion process, and the human visual system (HVS) is critical in algorithm development.  A prototypical example is the visual information fidelity (VIF) measure~\citep{sheikh2006image}. 
      
      \item {\it Learning-based methods} learn a metric from a training set of images and corresponding perceptual distances using supervised machine learning methods. By leveraging the power of DNNs, these methods have achieved state-of-the-art performance on existing image quality databases~\citep{bosse2018deep,prashnani2018pieapp}. But given the high dimensionality of the input space (\ie,  millions of pixels), these methods are prone to overfitting the limited available data. Strategies that compensate for the insufficiency of labeled training data include building on pre-trained networks~\citep{zhang2018unreasonable,ding2020dists}, training on local image patches \citep{bosse2018deep}, and combining multiple IQA databases \citep{zhang2019learning}.
      
      \item {\it Fusion-based methods} combine existing IQA methods to build a ``super-evaluator'' that exploits the diversity and complementarity of their constituent methods (analogous to ``boosting'' methods in machine learning). Fusion combinations can be determined empirically~\citep{ye2014beyond} or learned from data~\citep{liu2012imageq,ma2019blind}.
      Some methods incorporate deterministic or statistical image priors to regularize an IQA measure~\citep{jordan1881,ulyanov2018deep}. Since such regularizers can be seen as a form of no-reference IQA measures~\citep{wang2011reduced}, we also view these as fusion solutions.
  \end{itemize}
  
  
  \begin{figure*}
    \centering
      \subfloat[Initialization]{\includegraphics[height=0.16\linewidth]{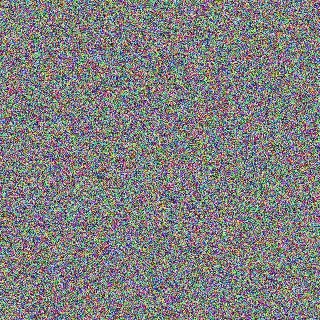}} \hskip0.2em
      \subfloat[MS-SSIM]{\includegraphics[height=0.16\linewidth]{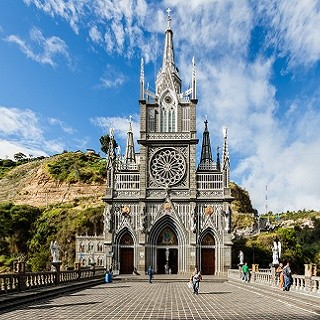}} \hskip0.2em
      \subfloat[IFC]{\includegraphics[height=0.16\linewidth]{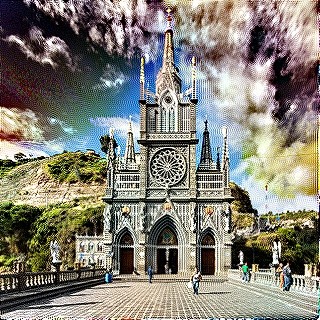}} \hskip0.2em
      \subfloat[VIF]{\includegraphics[height=0.16\linewidth]{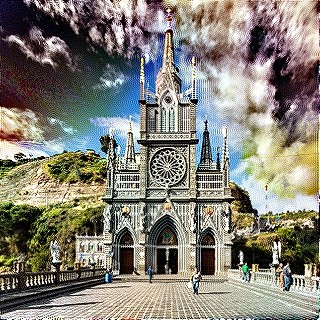}} \hskip0.2em
      \subfloat[CW-SSIM]{\includegraphics[height=0.16\linewidth]{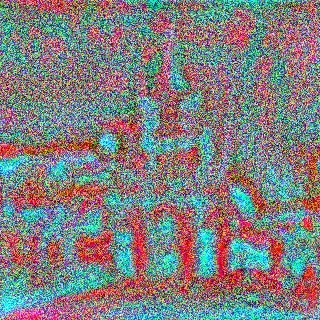}} \hskip0.2em
      \subfloat[MAD]{\includegraphics[height=0.16\linewidth]{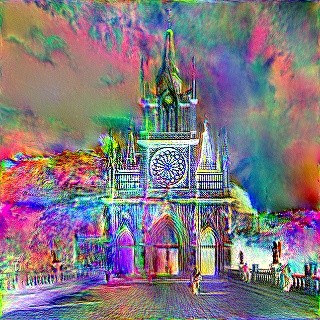}} \\\vspace{-1em}
      \subfloat[FSIM]{\includegraphics[height=0.16\linewidth]{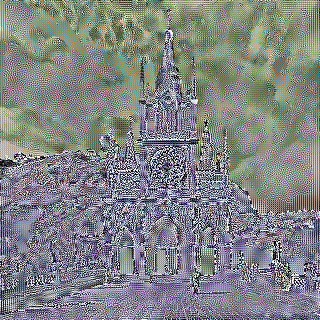}} \hskip0.2em  
      \subfloat[SFF]{\includegraphics[height=0.16\linewidth]{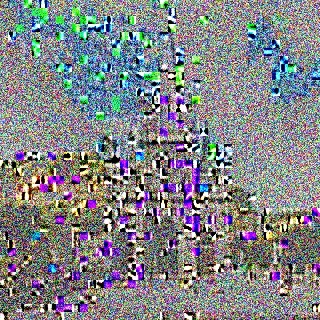}} \hskip0.2em  
      \subfloat[PAMSE]{\includegraphics[height=0.16\linewidth]{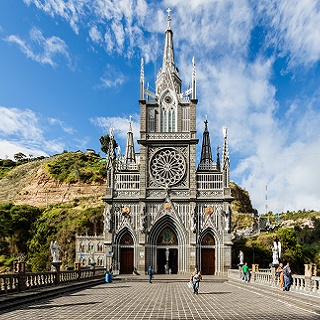}} \hskip0.2em   
      \subfloat[GMSD]{\includegraphics[height=0.16\linewidth]{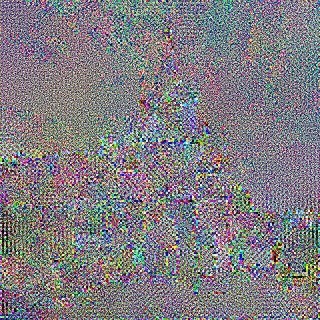}} \hskip0.2em 
      \subfloat[VSI]{\includegraphics[height=0.16\linewidth]{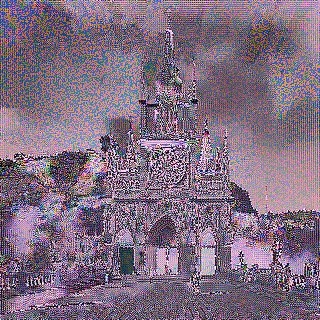}} \hskip0.2em
      \subfloat[MCSD]{\includegraphics[height=0.16\linewidth]{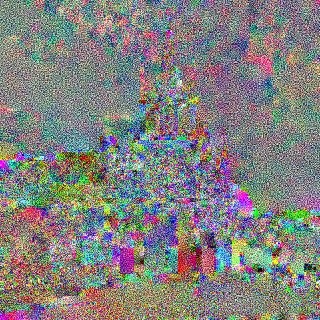}} \\\vspace{-1em}
      \subfloat[NLPD]{\includegraphics[height=0.16\linewidth]{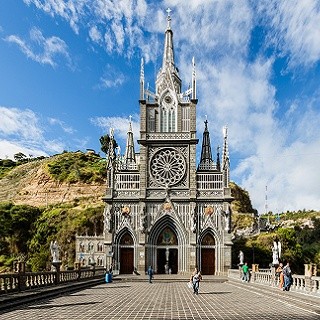}} \hskip0.2em
      \subfloat[GTI-CNN]{\includegraphics[height=0.16\linewidth]{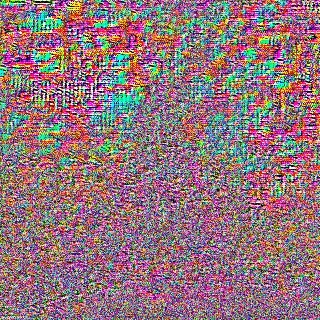}} \hskip0.2em
      \subfloat[DeepIQA]{\includegraphics[height=0.16\linewidth]{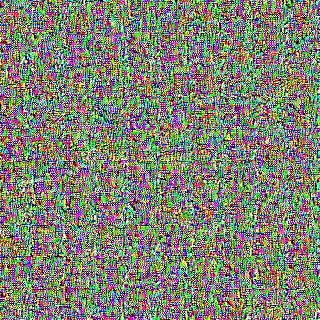}} \hskip0.2em
      \subfloat[PieAPP]{\includegraphics[height=0.16\linewidth]{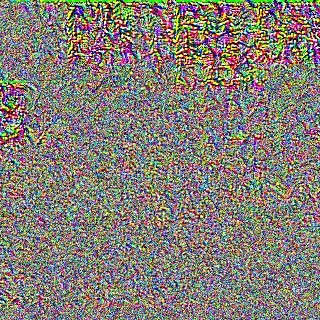}} \hskip0.2em
      \subfloat[LPIPS]{\includegraphics[height=0.16\linewidth]{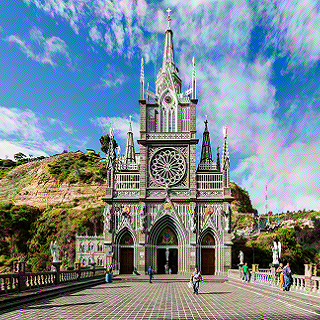}} \hskip0.2em
      \subfloat[DISTS]{\includegraphics[height=0.16\linewidth]{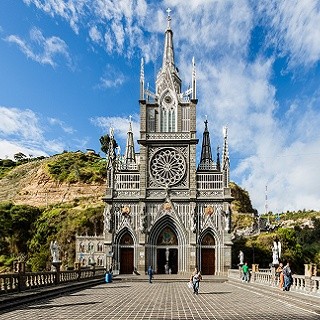}} \\ \vspace{-.2em}
    \caption{Reference image recovery test. Starting from (a) a white Gaussian noise image, we recover images by optimizing the predicted quality relative to a reference image, using different IQA models (b)-(r).} 
    \label{fig:toy1}
  \end{figure*}

  \section{Screening of Full-Reference IQA Models for Perceptual Optimization}\label{sec:experiments}
  We used a na\"{i}ve task to demonstrate the issues encountered when using IQA models in gradient-based perceptual optimization. This task also allows us to pre-screen existing models, and to motivate the design of experiments used in subsequent comparisons.  
  
  \subsection{Reference Image Recovery}\label{sec:toy}
  Given a reference (undistorted) image $x$ and an initial image $y_0$, we aimed to recover $x$ by numerically optimizing
  \begin{align}
      y^\star =\mathop{\arg\min}_{y} D(x,y) ,
  \end{align}
  where $D$ denotes a full-reference IQA measure with a lower score indicating higher predicted quality, and $y^\star$ is the recovered image. For example, if $D$ is MSE, the (trivial) analytical solution is $y^\star = x$, indicating full recoverability. The majority of current IQA models are continuous and differentiable, and solutions must be obtained numerically using gradient-based iterative solvers. We considered an initial set of $17$ methods, which we believe cover the full spectrum of full-reference IQA methods. These include three error visibility methods -  MAD~\citep{larson:011006}, PAMSE~\citep{xue2013perceptual} and NLPD~\citep{laparra2016perceptual}, seven structural similarity methods -  MS-SSIM~\citep{wang2003multiscale}, CW-SSIM~\citep{wang2005translation}, FSIM~\citep{zhang2011fsim}, SFF~\citep{chang2013sparse}, GMSD~\citep{xue2014gradient} and VSI~\citep{zhang2014vsi}, MCSD~\citep{wang2016multiscale}, two information-theoretical methods -  IFC~\citep{sheikh2005information} and VIF~\citep{sheikh2006image}, and five DNN methods -  GTI-CNN~\citep{ma2018geometric}, DeepIQA \citep{bosse2018deep}, PieAPP~\citep{prashnani2018pieapp}, LPIPS~\citep{zhang2018unreasonable} and DISTS~\citep{ding2020dists}. As this paper focuses on the perceptual optimization performance of individual IQA measures, fusion-based methods are not included.
  
  \begin{figure*}
    \centering
      \subfloat[Initialization]{\includegraphics[height=0.16\linewidth]{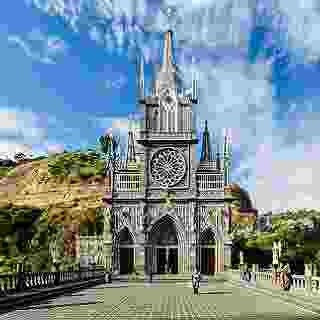}} \hskip0.2em
      \subfloat[MS-SSIM]{\includegraphics[height=0.16\linewidth]{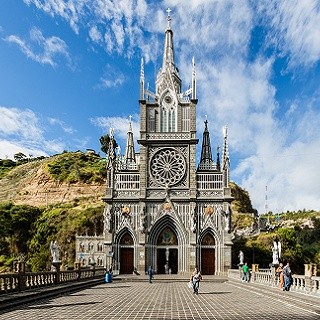}} \hskip0.2em
      \subfloat[IFC]{\includegraphics[height=0.16\linewidth]{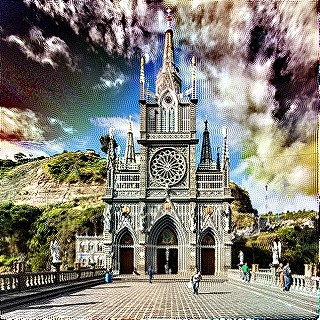}} \hskip0.2em
      \subfloat[VIF]{\includegraphics[height=0.16\linewidth]{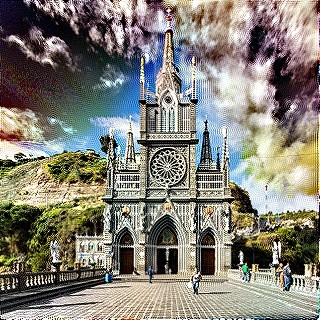}} \hskip0.2em
      \subfloat[CW-SSIM]{\includegraphics[height=0.16\linewidth]{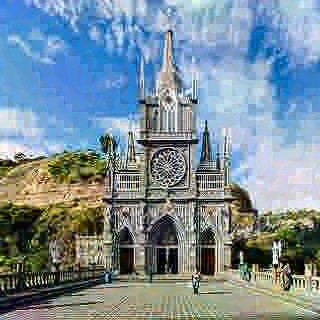}} \hskip0.2em
      \subfloat[MAD]{\includegraphics[height=0.16\linewidth]{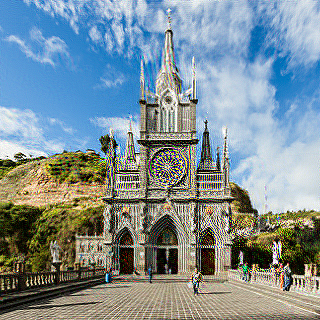}} \\\vspace{-1em}
      \subfloat[FSIM]{\includegraphics[height=0.16\linewidth]{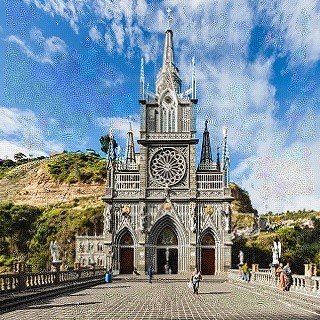}} \hskip0.2em  
      \subfloat[SFF]{\includegraphics[height=0.16\linewidth]{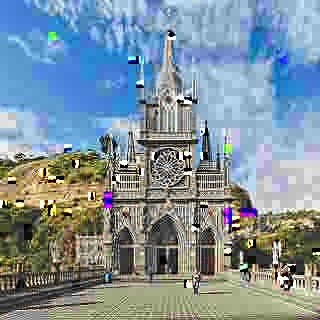}} \hskip0.2em  
      \subfloat[PAMSE]{\includegraphics[height=0.16\linewidth]{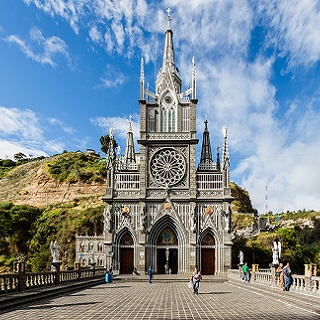}} \hskip0.2em   
      \subfloat[GMSD]{\includegraphics[height=0.16\linewidth]{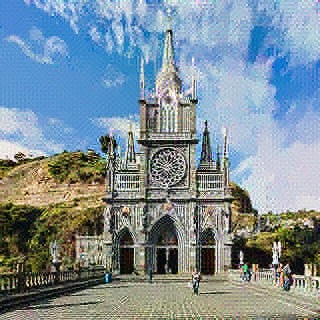}} \hskip0.2em 
      \subfloat[VSI]{\includegraphics[height=0.16\linewidth]{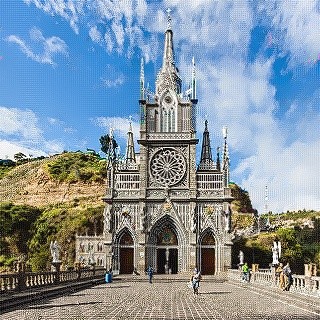}} \hskip0.2em
      \subfloat[MCSD]{\includegraphics[height=0.16\linewidth]{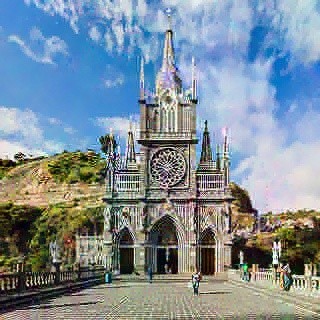}} \\\vspace{-1em}
      \subfloat[NLPD]{\includegraphics[height=0.16\linewidth]{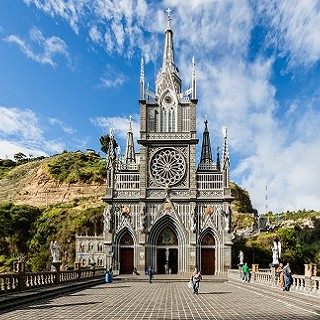}} \hskip0.2em
      \subfloat[GTI-CNN]{\includegraphics[height=0.16\linewidth]{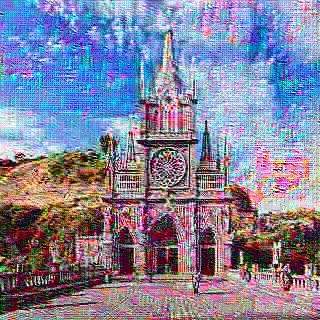}} \hskip0.2em
      \subfloat[DeepIQA]{\includegraphics[height=0.16\linewidth]{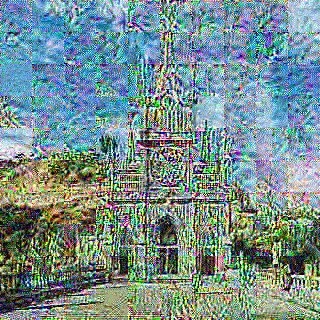}} \hskip0.2em
      \subfloat[PieAPP]{\includegraphics[height=0.16\linewidth]{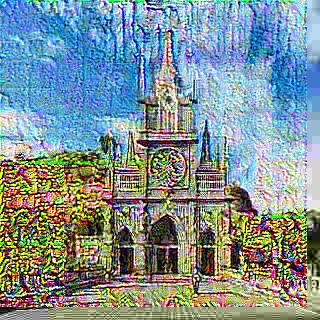}} \hskip0.2em
      \subfloat[LPIPS]{\includegraphics[height=0.16\linewidth]{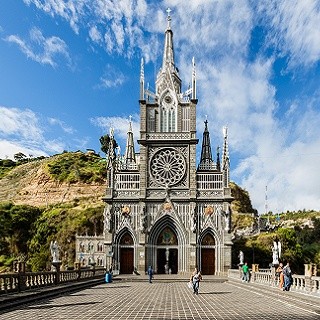}} \hskip0.2em
      \subfloat[DISTS]{\includegraphics[height=0.16\linewidth]{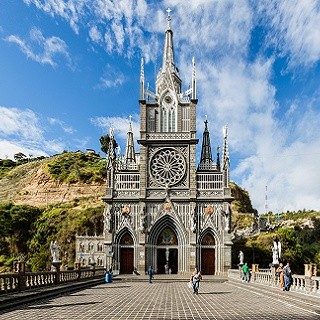}} \\ \vspace{-.2em}
    \caption{Reference image recovery test. Staring from  (a) a JPEG compressed version of a reference image, we recover images by optimizing the predicted quality relative to the reference image, using different IQA models (b)-(r).
    } 
    \label{fig:toy3}
  \end{figure*}
  
  Figures~\ref{fig:toy1} and~\ref{fig:toy3} show recovery results from two different initializations - a white Gaussian noise image and a JPEG-compressed version of a reference image, respectively. For all IQA methods, the optimization converges to a final image with a substantially better score than that of the initial image. Models based on injective mappings such as MS-SSIM, PAMSE, NLPD and DISTS are 
  able to recover the reference image (although the rate of convergence may depend on the choice of initial image). Many of the remaining IQA models generate a final image with worse visual quality than that of the initial image (e.g., compare Fig. \ref{fig:toy3} (a) with (o) or (p)), often with noticeable model-dependent artifacts.  This is because these methods rely on surjective mapping functions to transform the images to a reduced ``perceptual'' space for quality computation. 
  For example,  GTI-CNN~\citep{ma2018geometric} uses a surjective DNN with four stages of convolution, subsampling, and halfwave rectification.  The resulting undercomplete representation is optimized for geometric transformation invariance, at the cost of significant  information loss. The examples demonstrate that preservation of some aspects of this lost information is important for perceptual quality.  Similar arguments can be applied to other surjective DNN-based IQA models, such as  DeepIQA~\citep{bosse2018deep} and  PieAPP~\citep{prashnani2018pieapp}.  Generally, optimization guided by the surjective models ``recovers'' more structures when initialized with the JPEG image (which provides roughly correct local luminances), as compared to initialization with purely white Gaussian noise. 

  \subsection{IQA Model Selection}
  The reference image recovery test results were used to pre-screen the initial set of IQA models, excluding those that perform poorly (due to surjectivity). In addition, we excluded models with similar designs. This process yielded $11$ full-reference IQA models to be compared in our human subject evaluations:
  
  \begin{enumerate}
  \item MAE, the Mean Absolute Error ($\ell_1$-norm) of pixel values, has been frequently adopted in optimization, despite its poor perceptual relevance. MAE has been shown to consistently outperform MSE ($\ell_2$-norm) in image restoration tasks \citep{zhao2016loss}.
  
  \item MS-SSIM~\citep{wang2003multiscale}, the Multi-Scale extension of the SSIM index~\citep{wang2004image}, provides more ﬂexibility than single-scale SSIM, allowing for a wider range of viewing distances. 
  It decomposes the input images into Gaussian pyramids \citep{burt1983laplacian}, and computes contrast and structure similarities at each scale and luminance similarity at the coarsest scale only.
  MS-SSIM has become a standard ``perceptual'' quality measure, and has been used to guide the design of DNN-based image super-resolution \citep{zhao2016loss,snell2017learning} and compression \citep{ball2018variational} algorithms.
  
  \item VIF~\citep{sheikh2006image}, the Visual Information Fidelity measure, quantifies how much information from the reference image is preserved in the distorted image. A Gaussian scale mixture \citep{portilla2003gsm} is used as a source model to summarize natural image statistics, and mutual information is estimated assuming only signal attenuation and additive noise perturbations. A distinct property of VIF relative to other IQA models is that it can handle cases in which the ``distorted'' image is visually superior to the reference~\citep{wang2015patch}. 
  
  \item CW-SSIM~\citep{wang2005translation}, the Complex Wavelet SSIM index, is designed to be robust to small geometric distortions such as translation and rotation. The construction allows for consistent local phase shifts of  wavelet coefficients, which preserves image features. CW-SSIM addresses a common limitation of IQA methods that require precise spatial registration of the reference and distorted images. 
  
  \item MAD~\citep{larson:011006}, the Most Apparent Distortion measure, explicitly models adaptive strategies of the HVS. Specifically, a detection-based strategy considering local luminance and contrast masking is employed for near-threshold distortions, and an appearance-based strategy involving local spatial-frequency statistics is activated for supra-threshold distortions. The two strategies are combined by a weighted geometric mean, where the weight is determined based on the amount of distortion. 
  
  \item FSIM~\citep{zhang2011fsim}, the Feature SIMilarity index, assumes that HVS understands an image mainly according to its low-level features. It computes quality estimates based on phase congruency~\citep{kovesi1999image} as the primary feature, and incorporates the gradient magnitude as the complementary feature. Moreover, the phase congruency component serves as a local weighting 
  factor to derive an overall quality score. FSIM also supplies a color version by making quality measurements from chromatic components.
  
  \item GMSD~\citep{xue2014gradient}, the Gradient Magnitude Similarity Deviation, focuses on computational efficiency of quality prediction, by simply computing pixel-wise gradient magnitude similarity followed by standard deviation (std) pooling. This pooling strategy is, however, problematic because an image with large but constant local distortion yields an std of zero (indicating the best predicted quality).
  
  \item VSI~\citep{zhang2014vsi}, the Visual Saliency Induced quality index, assumes that the change of salient regions due to image degradation is closely related to  the change of visual quality. The saliency map is used not only as a quality feature, but also as a weighting function to characterize the importance of a local region. By combining saliency magnitude, gradient magnitude and 
  chromatic features, VSI demonstrates good quality prediction performance, especially for localized distortions, such as local patch substitution~\citep{Ponomarenko201557}.
  
  \item NLPD~\citep{laparra2016perceptual}, the Normalized Laplacian Pyramid Distance, mimics the nonlinear transformations of the early visual system: local luminance
  subtraction and local gain control, and combines these values using weighted $\ell_p$-norms. The parameters are optimized to minimize the representation redundancies, instead of matching human judgments. NLPD has been successfully employed to optimize image rendering algorithms~\citep{ma2015high,Laparra:17}, where the input
  reference image has a much higher dynamic range than that of the display.  It has also been used to optimize a compression system~\citep{balle2016end}.
  
  \item LPIPS~\citep{zhang2018unreasonable}, the Learned Perceptual Image Patch Similarity model, computes the Euclidean distance between deep representations of two images. The authors showed that feature maps of different DNN architectures have ``reasonable'' effectiveness in accounting for human perception of image quality. As LPIPS has many different configurations, we chose the default one based on the VGG network~\citep{Simonyan14c} with the weights learned from the BAPPS dataset~\citep{zhang2018unreasonable}. VGG-based LPIPS can be seen as a generalization of the ``perceptual loss'' \citep{johnson2016perceptual}, which computes the Euclidean distance on convolution responses from one stage of VGG.
  
  \item DISTS~\citep{ding2020dists}, the Deep Image Structure and Texture Similarity metric, is explicitly designed to tolerate texture resampling (\eg, replacing one patch of grass with another).  DISTS is based on an injective mapping function built from a variant of the VGG network, and combines SSIM-like  structure and texture similarity measurements between corresponding feature maps of the two images. It is sensitive to structural distortions but at the same time robust to texture resampling and modest geometric transformations.
  \end{enumerate}
  
  We re-implemented all $11$ of these models using PyTorch\footnote{\url{https://pytorch.org}}, and verified that our code could reproduce the published performance results for each model on the LIVE~\citep{LIVE}, CSIQ~\citep{larson:011006}, and TID2013~\citep{Ponomarenko201557} databases (see Table~\ref{tab:re_srcc} in Appendix~\ref{ap:assess}).
  We modified grayscale-only models to accept color images, by computing scores on RGB channels separately and averaging them to obtain an overall quality estimate. 
  
  \begin{figure*}
    \centering
      \includegraphics[height=0.22\linewidth]{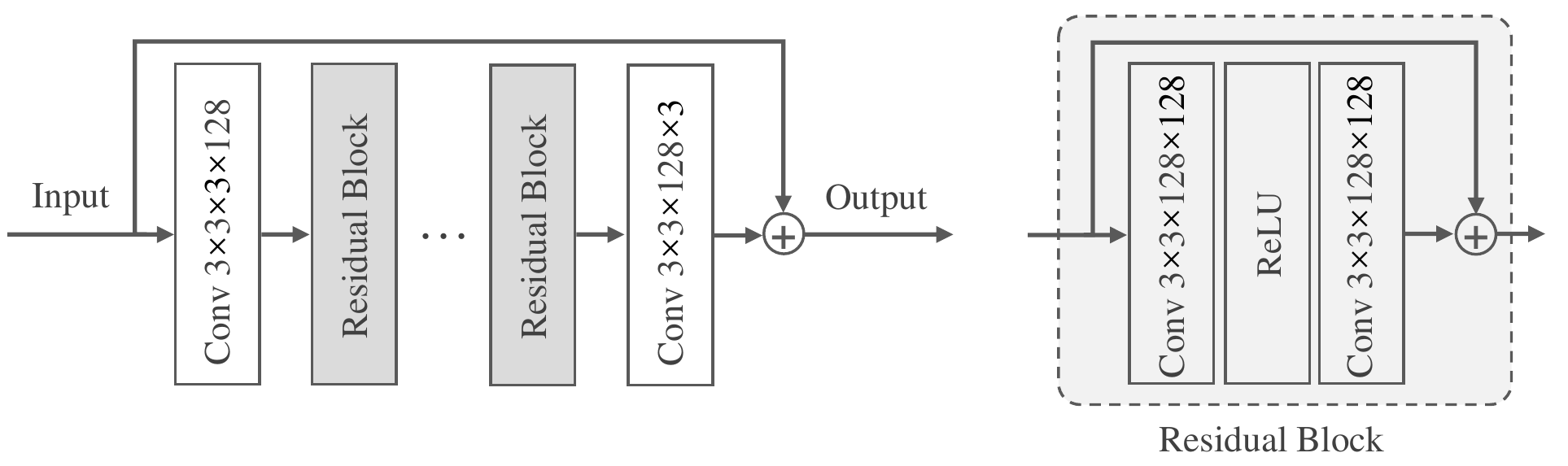}
    \caption{Network architecture used for denoising and deblurring. In addition to initial and final convolutional blocks, it contains $16$ residual blocks, each consisting of two convolutions and a halfwave rectifier (ReLU). Conv $h\times w\times c_{\text{in}}\times c_{\text{out}}$ indicates affine convolution with filter size $h\times w$, over $c_{\text{in}}$ input channels, producing  $c_{\text{out}}$ output channels.}  
    \label{fig:net_denoise}
  \end{figure*}

  \begin{figure*}
    \centering
      \includegraphics[height=0.18\linewidth]{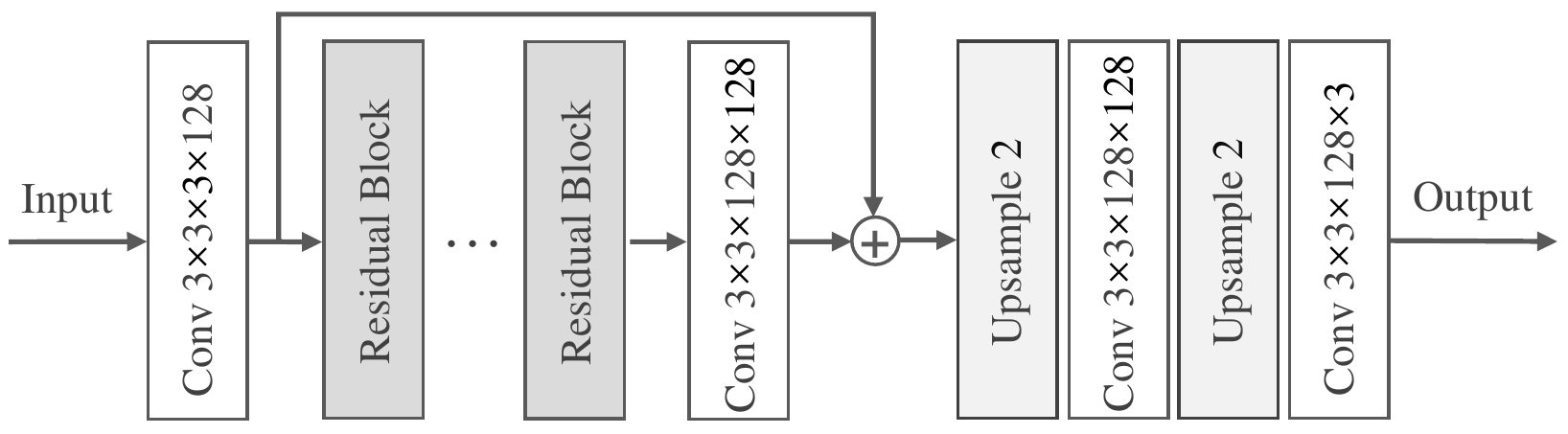}
    \caption{Network architecture used for super-resolution, containing $16$ residual blocks followed by two upsampling modules, each composed of an upsampler (factor of $2$, using nearest-neighbor interpolation) and a convolution.}  
    \label{fig:net_sr}
  \end{figure*}

  \begin{figure*}
    \centering
      \includegraphics[height=0.22\linewidth]{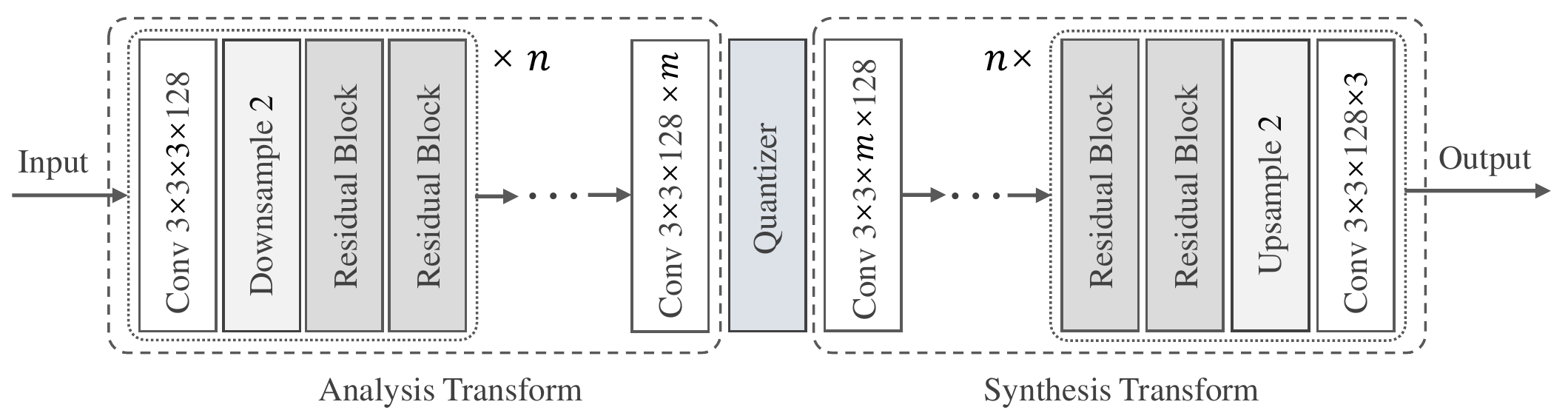}
    \caption{Network architecture used for lossy image compression, which includes an analysis transformation $f_a$, a quantizer $Q$, and a synthesis transformation $f_s$. $f_a$ is comprised of $n$ blocks, each with a convolution and downsampling (stride) by $2$ followed by two residual blocks. After the last block, another convolution layer with $m$ filters is added to produce the internal code representation, the values of which are then quantized by $Q$. $f_s$ consists of a cascade that is mirror-symmetric to $f_a$, with nearest-neighbor interpolation used to upsample the feature maps. }  
    \label{fig:net_comp}
  \end{figure*}
  
  \section{Perceptual Optimization of Standard Image Processing Tasks}
  We used each of the $11$ full-reference IQA models to guide the learning of  DNNs to solve four low-level vision tasks:
  \begin{itemize}
      \item image denoising,
      \item blind image deblurring,
      \item single image super-resolution,
      \item lossy image compression.
  \end{itemize} 
  The parameters of each network are optimized to minimize an IQA measure over a database of corrupted and original image pairs via stochastic gradient descent. 
  Implementations of all IQA models, as well as the DNNs for the four tasks, are available at \url{https://github.com/dingkeyan93/IQA-optimization}.
  
  \subsection{Image Denoising} \label{subsec:denoise}
  Image denoising is a core application of classical image processing, and also plays an essential role in testing prior models of natural images. In its simplest form, one aims to recover an unknown clean image $x\in\mathbb{R}^{N}$ from an observed image $y$ that has been corrupted by additive white Gaussian noise $n$ of known variance $\sigma^2$, \ie, $y=x+n$. Denoising algorithms can be roughly classified into spatial domain methods (\eg, Wiener filter~\citep{wiener1950}, bilateral filter~\citep{tomasi1998bilateral} and collaborative filtering~\citep{dabov2007image}), and wavelet transform methods~\citep{donoho1995shrinkage,simoncelli1996noise,portilla2003gsm}. Adaptive sparsifying transforms \citep{elad2006image} and variants of nonlinear shrinkage functions have also been directly learned from natural image data~\citep{hel2008discriminative,Raphan2008Optimal}.
  In recent years, purely data-driven models based on DNNs have achieved state-of-the-art levels of performance~\citep{zhang2017beyond}. 
  
  Here, we constructed a simplified DNN, shown in  Fig.~\ref{fig:net_denoise}, inspired by the EDSR network~\citep{lim2017enhanced}.  The network was trained to estimate the noise (which is then subtracted from the observation to yield a denoised image), by minimizing a loss function defined as
  \begin{align}
      \ell(\phi)= D\left(y-f_\phi(y), x\right),
    \label{eq:loss_denoise}
  \end{align}
  where $D$ is an IQA measure and $f_\phi : \mathbb{R}^{N} \mapsto \mathbb{R}^{N}$ is the mapping of the DNN, parameterized by vector $\phi$. 
  
  \subsection{Blind Image Deblurring} \label{subsec:deblur}
  The goal of image deblurring is to restore a sharp image $x$ from a blurry observation $y$, which can occur due to defocus and motion of the camera, and motion of objects in a scene. The observation process is usually described by 
  \begin{align}
      y = Kx + n,
  \end{align}
  where $K\in\mathbb{R}^{N\times N}$ denotes a spatially-varying linear kernel. Blind deblurring refers to the problem in which the blur kernel is unknown. Most early methods, \eg, the classical Lucy-Richardson algorithm~\citep{richardson1972bayesian,lucy1974iterative}, focused on non-blind deblurring where the blur kernel is assumed known.  Successful blind deblurring methods, such as ~\citep{Freeman2006Removing,pan2016blind}, rely heavily on statistical priors of natural images and geometric priors of blur kernels. With the success of deep learning, many DNN-based approaches \citep{tao2018scale,kupyn2018deblurgan} attempt to directly learn the mapping function for blind deblurring without explicitly estimating the blur kernel. Here we also adopted this ``kernel-free'' approach to train a DNN for image deblurring in an end-to-end fashion. We employed the same network architecture used in denoising (see Fig.~\ref{fig:net_denoise}) with the same loss function (Eq.~\eqref{eq:loss_denoise}).

  \subsection{Single Image Super-Resolution} \label{subsec:sr}
  Single image super-resolution aims to enhance the resolution and quality of a low-resolution image, which can be modelled by
  \begin{align}
      y = PKx +n,
  \end{align}
  where $P$ denotes downsampling by
  a factor of $\beta$. This is an ill-posed problem, as downsampling is a projection onto a lower-dimensional subspace, and its solution must rely on some form of regularization or prior model. Early attempts exploited sampling theory~\citep{li2001new} or natural image statistics~\citep{sun2008image}. Later methods focused on learning mapping functions between the low-resolution and high-resolution images through sparse coding~\citep{yang2010image}, locally linear regression~\citep{timofte2013anchored}, self-exemplars~\citep{Huang2015Single}, etc.
  Since 2014, DNN-based methods have come to dominate this field as well~\citep{dong2014learning}. An efficient method of constructing a DNN-based mapping is to first extract features from the low-resolution input and then upscale them with sub-pixel convolution~\citep{shi2016real,lim2017enhanced}. Here, we followed this method in constructing a DNN-based function $f:\mathbb{R}^{\left\lfloor\frac{N}{\beta^2}\right\rfloor}$ $\mapsto \mathbb{R}^N$, with architecture specified  in  Fig.~\ref{fig:net_sr}. The loss is specified by 
  \begin{align}
      \ell(\phi)= D\left(f_\phi(y), x\right).
    \label{eq:loss_sr}
  \end{align}

  \subsection{Lossy Image Compression} \label{subsec:compression}
  Data compression involves finding a more compact data representation from which the original image can be reconstructed. Compression can be either lossless or lossy. Here we followed a prevailing scheme in lossy image compression - transform coding, which consists of  transformation, quantization, and entropy coding. 
  Traditional image compression methods (\eg, the most widely used standard - JPEG) used a fixed linear transform for all bit rates. More recently, many researchers have demonstrated the visual benefits of nonlinear transforms, especially DNN-based learnable ones that are capable of adapting their parameters to different bitrate budgets.
  In this paper, we constructed two DNNs for analysis and synthesis transforms, respectively, as shown in Fig.~\ref{fig:net_comp}. The analysis transform $f_a$ maps the image to a latent feature vector $z$, whose values are then quantized to $L$ levels with the centers being $\{c_1, \ldots, c_L\}$, where $c_i \in\mathbb{R}$ for $i=1,\ldots, L$. This  quantized representation $\bar{z} = Q(f_a(x))$,  is fed to the synthesis transform $f_s$ to reconstruct the compressed image: $y = f_s(\bar{z})$. The quantizer has zero gradients almost everywhere (and infinite gradients at the transitions), which prevents training via gradient descent~\citep{Balle17a}. Hence, we used a soft differentiable approximation~\citep{mentzer2018conditional}
  \begin{align}
  \bar{z}_{i} = Q(z_{i})=\sum_{j=1}^{L} \frac{\exp \left(-s(z_{i}-c_{j})^2\right)}{\sum_{k=1}^{L} \exp \left(-s(z_{i}-c_{k})^2\right)} c_{j}
  \label{eq:soft_q}
  \end{align}
  to backpropagate gradients during training, where the scale parameter $s$ controls the degree to which $Q(\cdot)$ approximates quantization.
  
  In lossy image compression, the objective function is a weighted sum of two terms that quantify the coding cost and the reconstruction error, respectively:
  \begin{align}
  \ell = \lambda H[\bar{z}]+  \mathbb{E}[D(y, x)] .
  \label{eq:rd}
  \end{align}
  The first term is typically the entropy of the discrete codes $\bar{z}$, which provides a lower bound on the bitrate for transmitting the quantized coefficients~\citep{Shannon1948}.  The second term is the distortion between the reconstructed image $y$ and the original image $x$, as quantified by the full-reference IQA model $D$. The Lagrange  multiplier $\lambda$ controls the rate-distortion trade-off. 
  Due to substantially different scales of IQA model values, $\lambda$ should be adjusted for each model in order to enable fair comparison at similar bitrates, an extremely time-intensive process. To avoid this, following~\cite{agustsson2019generative}, we set $\lambda=0$ in Eq.~\eqref{eq:rd}, and controlled an upper bound on bitrate 
  \begin{equation}
  H(\bar{z}) \leq \mathrm{dim}(\bar{z}) \log_{2}(L)
  \end{equation}
  by adjusting the architecture of $f_s$ (\ie, the dimension of $\bar{z}$) and the number of quantization levels $L$ in $Q$. This elimination of the entropy from the objective also means that we did not need to continually re-estimate the probability mass function $P(\bar{z})$, which varies with changes in the network parameters.
  The optimization objective in Eq.~\eqref{eq:rd} is reduced to
  \begin{equation}
      \ell(\phi,\psi)=  \mathbb{E}\left[D\bigg(f_{s,\psi}\Big(Q\big(f_{a,\phi}(x)\big)\Big), x\bigg)\right],
  \label{eq:loss_comp}
  \end{equation}
  where $\phi$ and $\psi$ are the parameters of $f_a$ and $f_s$, respectively. The expectation is approximated by averaging over mini-batches of training images.

  \section{Implementation Issues}
  \label{subsec:details}
  In this section, we describe in detail the training of our DNN-based  computational models for the four low-level vision tasks, and the subjective testing procedure used to collect human ratings of the optimized images. 
  
  \subsection{Model Training}
  For denoising, we fixed the  noise std  to $\sigma=50$ (relative to pixel values in the range $[0,255]$).
  For deblurring, we simulated various kernels with different motion patterns and blur levels as in~\cite{kupyn2018deblurgan}.
  For super-resolution, we generated low-resolution images by downsampling high-resolution images by a factor of $\beta = 4$ using bicubic interpolation. 
  For compression, we set the number of quantization levels to $L=2$ with centers $\{-1, 1\}$, the quantization scale parameter to $s=1$, the number of downsampling stages to $n=4$, and the number of output channels of $f_a$ to $m=64$. This  leads to a maximum of $\frac{H(\bar{z})}{W \times H} \leq \frac{W \times H}{2^4 \cdot 2^4} \cdot 64 \cdot \log _{2}(2) / (W \times H)=0.25$ bits per pixel (bpp). 
  
  \begin{figure}
    \centering
      \subfloat{\includegraphics[height=0.24\linewidth]{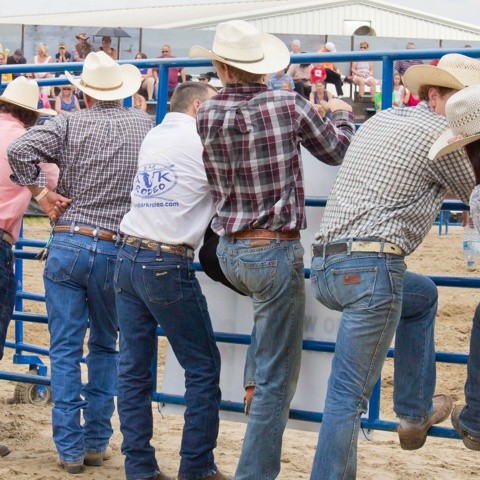}} \hskip0.2em
      \subfloat{\includegraphics[height=0.24\linewidth]{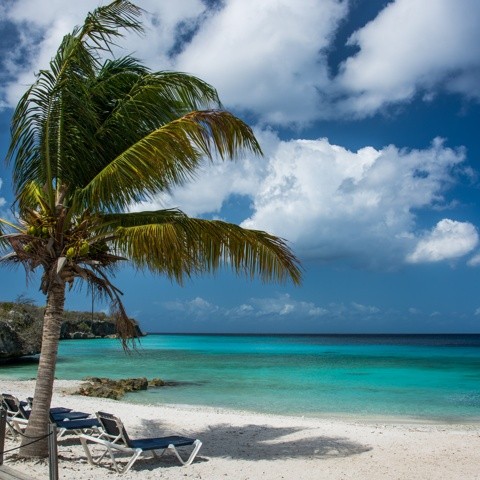}} \hskip0.2em
      \subfloat{\includegraphics[height=0.24\linewidth]{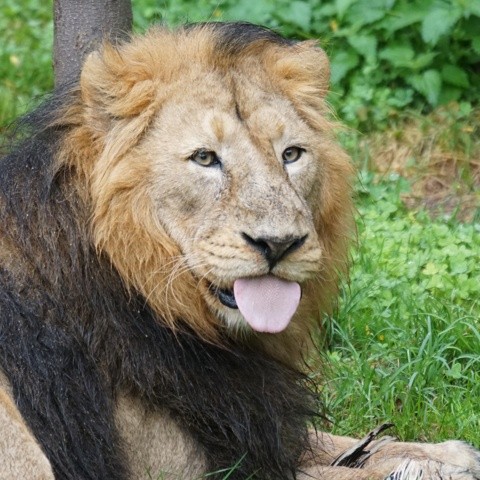}} \hskip0.2em
      \subfloat{\includegraphics[height=0.24\linewidth]{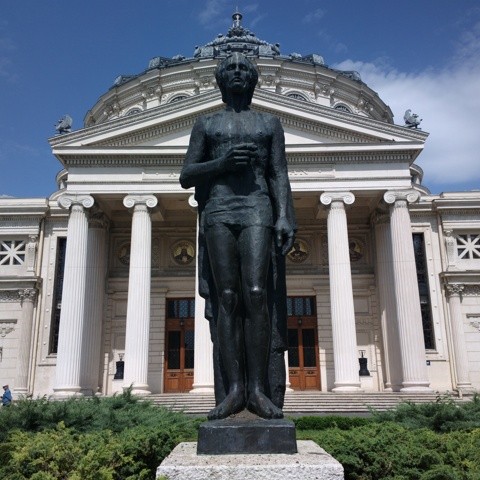}} \\ \vspace{-1em}
      \subfloat{\includegraphics[height=0.24\linewidth]{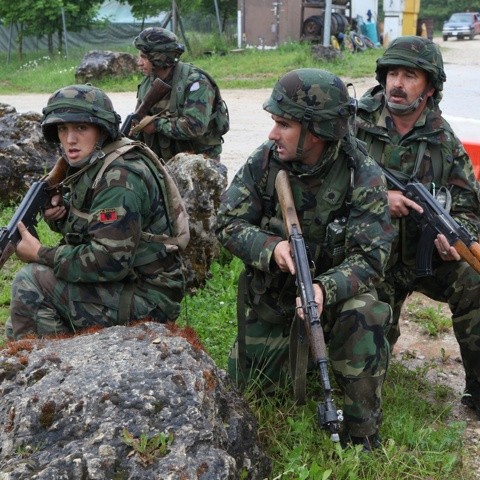}} \hskip0.2em
      \subfloat{\includegraphics[height=0.24\linewidth]{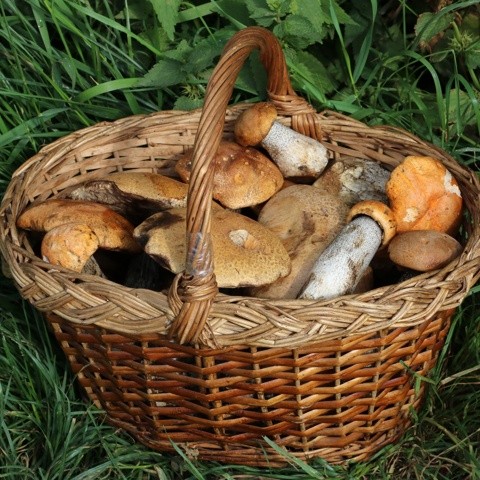}} \hskip0.2em 
      \subfloat{\includegraphics[height=0.24\linewidth]{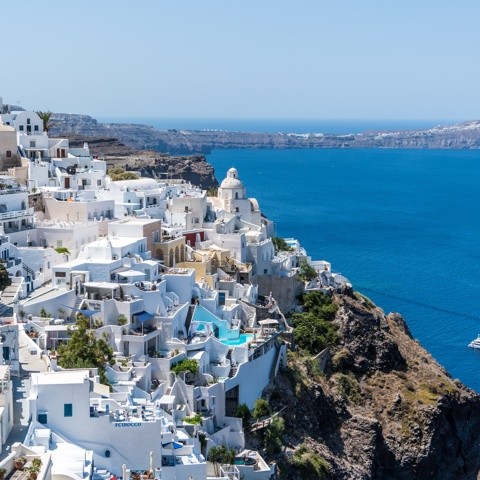}} \hskip0.2em
      \subfloat{\includegraphics[height=0.24\linewidth]{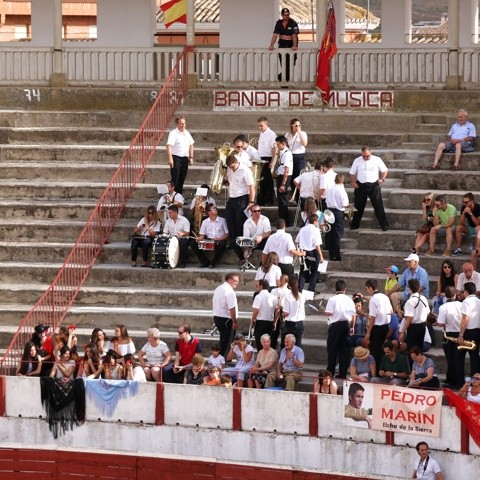}} \\  \vspace{-1em}
      \subfloat{\includegraphics[height=0.24\linewidth]{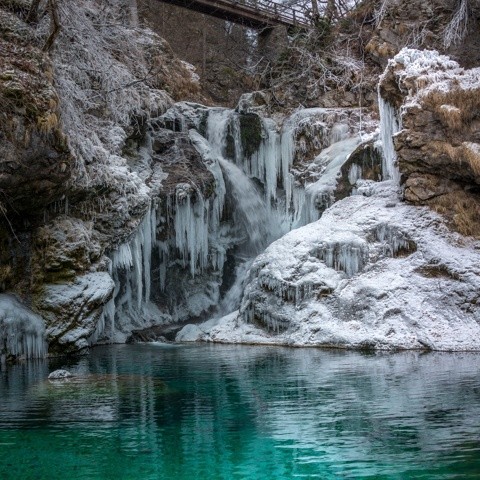}} \hskip0.2em
      \subfloat{\includegraphics[height=0.24\linewidth]{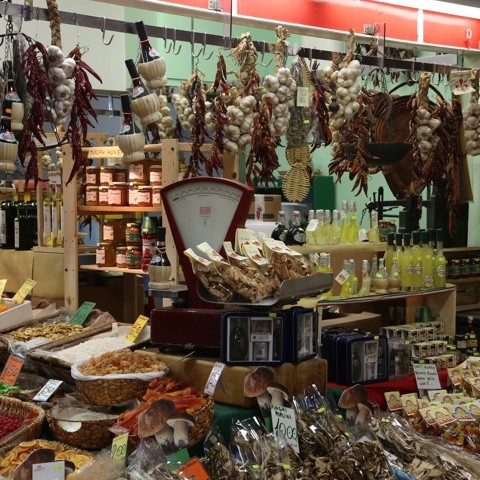}} \hskip0.2em
      \subfloat{\includegraphics[height=0.24\linewidth]{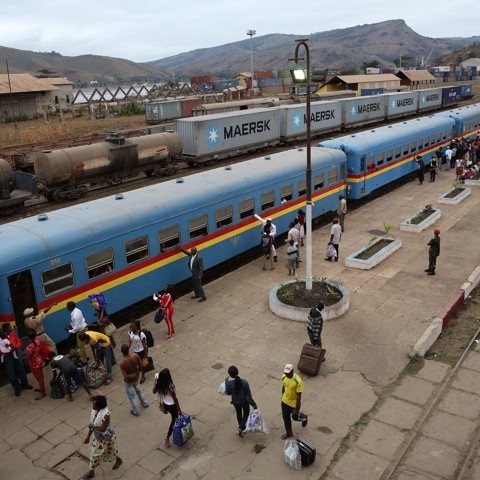}} \hskip0.2em
      \subfloat{\includegraphics[height=0.24\linewidth]{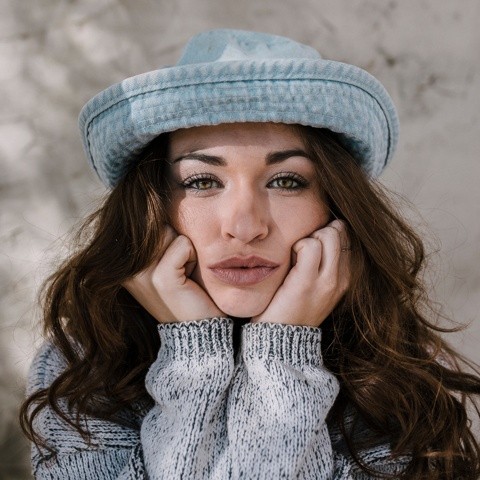}} \\  \vspace{-1em}
      \subfloat{\includegraphics[height=0.24\linewidth]{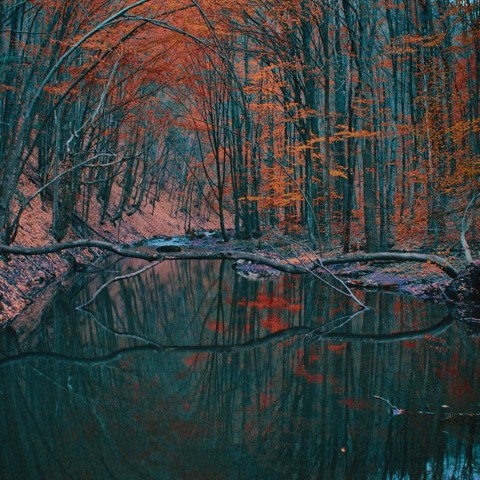}} \hskip0.2em
      \subfloat{\includegraphics[height=0.24\linewidth]{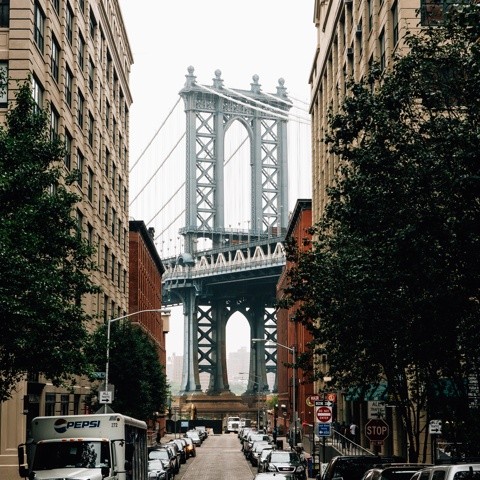}} \hskip0.2em
      \subfloat{\includegraphics[height=0.24\linewidth]{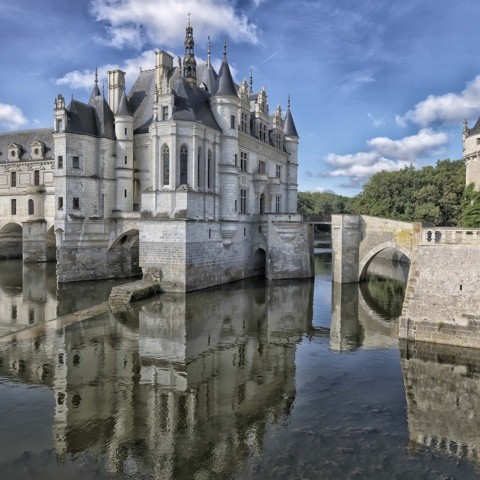}} \hskip0.2em
      \subfloat{\includegraphics[height=0.24\linewidth]{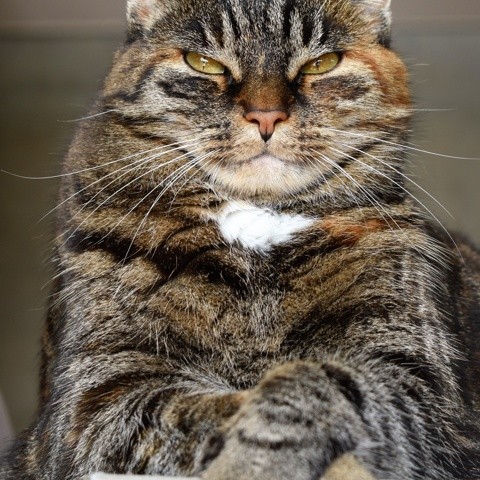}} \\  \vspace{-1em}
      \subfloat{\includegraphics[height=0.24\linewidth]{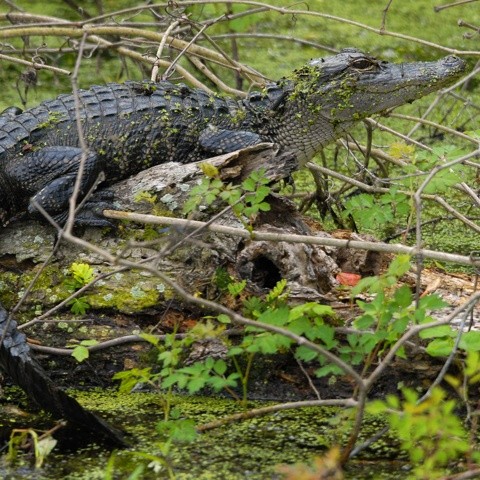}} \hskip0.2em
      \subfloat{\includegraphics[height=0.24\linewidth]{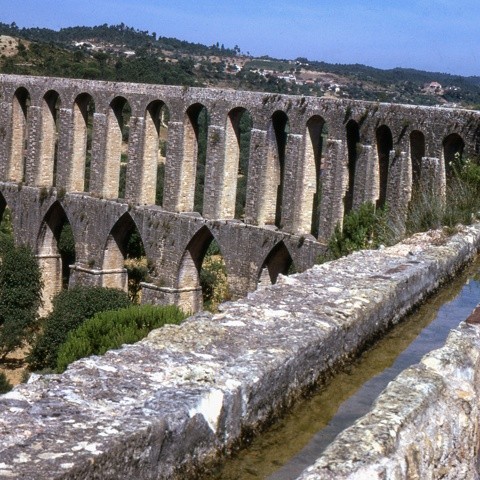}} \hskip0.2em
      \subfloat{\includegraphics[height=0.24\linewidth]{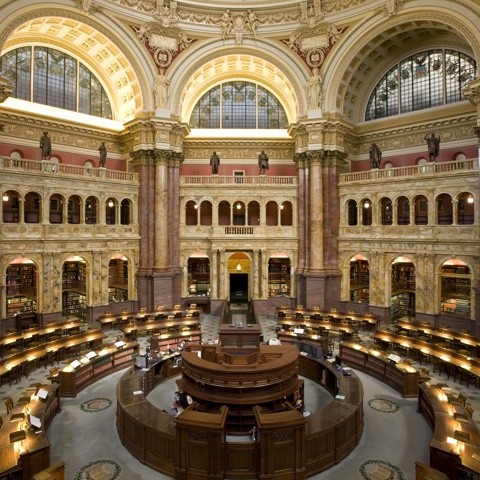}} \hskip0.2em
      \subfloat{\includegraphics[height=0.24\linewidth]{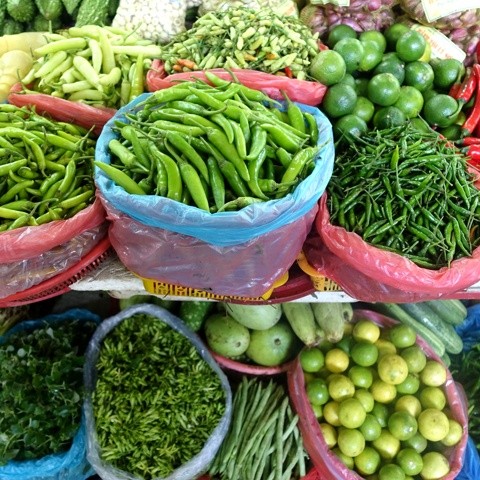}} 
    \caption{Test images (from the validation set of DIV2K) used in the subjective experiment.} 
    \label{fig:testing_imgs}
  \end{figure}

  \begin{figure*}
    \centering
      \subfloat[Denoising]{\includegraphics[height=0.35\linewidth]{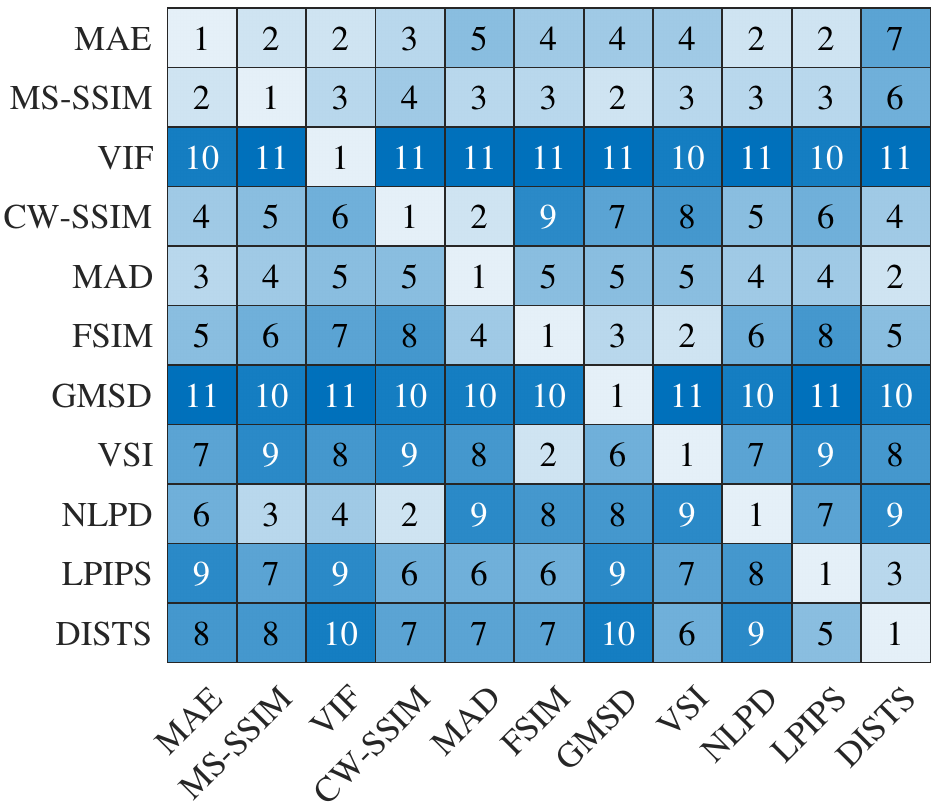}} \hskip2.em
      \subfloat[Deblurring]{\includegraphics[height=0.35\linewidth]{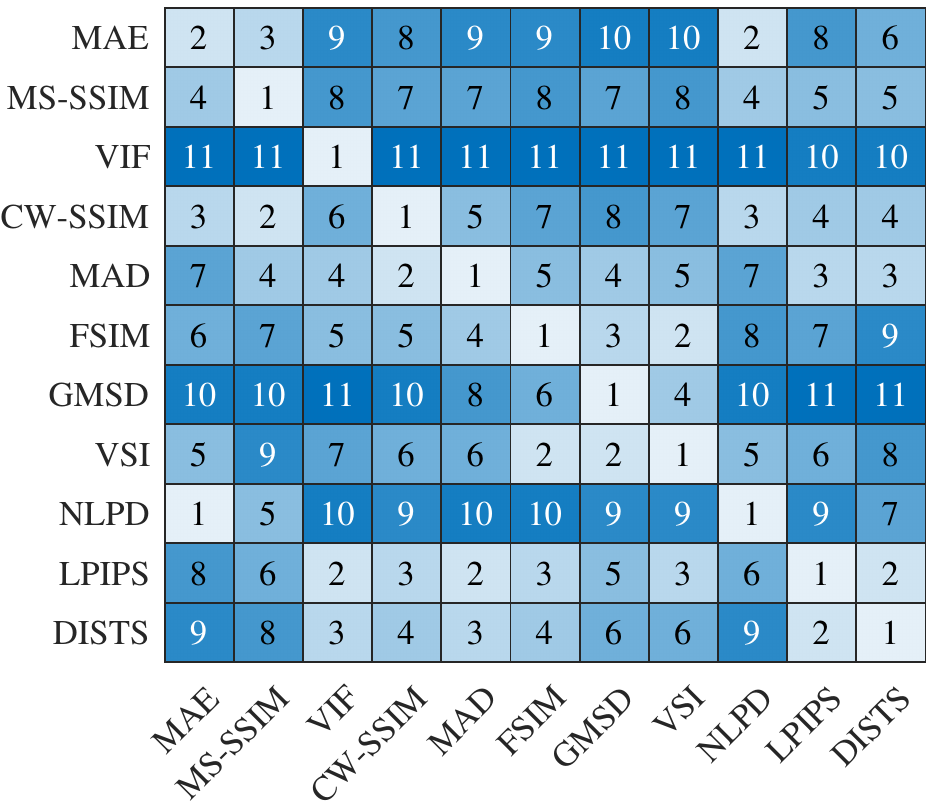}} \\
      \vspace{-0.5em}
      \subfloat[Super-resolution]{\includegraphics[height=0.35\linewidth]{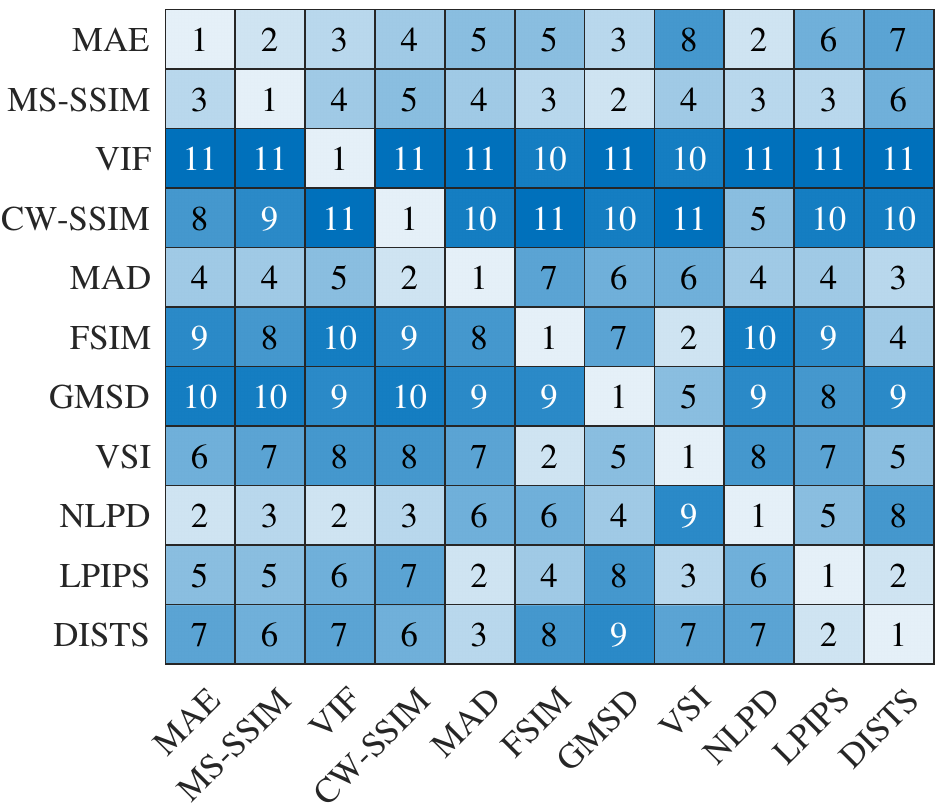}} \hskip2.em
      \subfloat[Compression]{\includegraphics[height=0.35\linewidth]{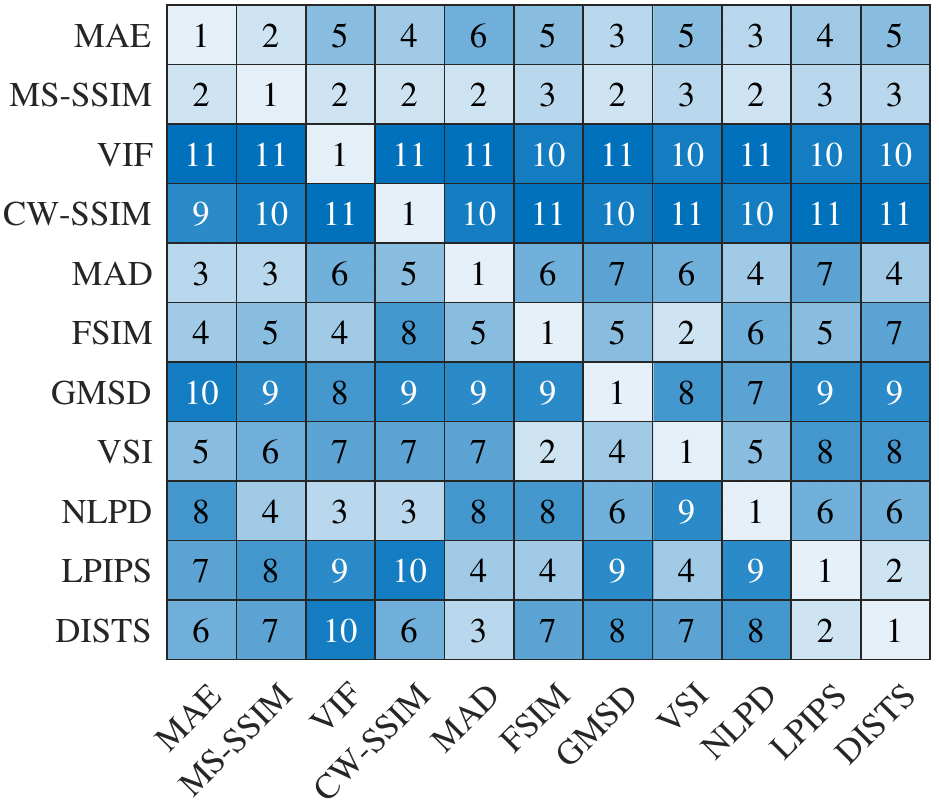}}
    \caption{Objective ranking of the final results in the four tasks. Vertical axis indicates IQA models used to train the networks, and horizontal axis indicates IQA models used to evaluate performance. The numbers of $1$ to $11$ indicate the rank order from the best to the worst.} 
    \label{fig:matrix}
  \end{figure*}

  We chose the $4,744$ high-quality images in the Waterloo Exploration Database~\citep{ma2016waterloo} as reference images. Training was performed in two stages. In the first stage, we pre-trained a network using MAE as the loss function for all four tasks \citep{wang2018esrgan}. In the second stage, we fine-tuned the network parameters by optimizing the desired IQA model. Pre-training brings several advantages. First, some IQA models are sensitive to initializations (\eg, CW-SSIM, MAD, FSIM, GMSD, and VSI) and pre-training yields more reasonable optimization results (also validated in the task of reference image recovery). Second, models that require backpropagating gradients through multiple stages of computation (e.g., LPIPS and DISTS) converge much faster. Third, it helps us to test whether the recently proposed IQA models lead to consistent perceptual gains on top of MAE, a special case of the simple $\ell_p$-norm distance.

  For each training stage of the four tasks, we used the Adam optimization package~\citep{kingma2014adam} with a mini-batch size of $16$ and an initial learning rate of $10^{-4}$, which decays linearly by a factor of $2$ for every $100$K iterations, and we set the maximum number of iterations to $500$K. We randomly extracted patches with the size of $192\times192\times3$ during training, and tested on $20$ independent images selected from the DIV2K validation set (see Fig.~\ref{fig:testing_imgs}). Training took roughly $1,000$ GPU hours (measured using an NVIDIA GTX 2080 device) for a total of $4\times 11 = 44$ models. Special treatments (\ie, gradient clipping and a smaller learning rate) were given to FSIM and VSI, otherwise their losses are difficult to converge according to our trials.
  
  Generally, it can be difficult to stabilize the training of DNNs to convergence, especially given that the gradients of different IQA models exhibit idiosyncratic behaviors. Fortunately, a simple criterion exists to test the validity of the optimization results: for a given low-level vision task, the DNN optimized for the IQA measure $D_i$ should produce the best result (averaged over an independent set of images) in terms of $D_i$ itself, when comparing to DNNs optimized for $\{D_j\}_{j\ne i}$. Fig.~\ref{fig:matrix} shows the ranking of results generated by networks optimized for each of the $11$ IQA models (corresponding to one column in one subfigure) on the DIV2K validation set~\citep{timofte2017ntire}, where $1$ and $11$ indicate the best and worst rankings, respectively. By inspecting the diagonal elements of the four matrices, we conclude that $43$ out of $44$ models satisfy the criterion, verifying the rationality of our training procedures.  The only exception is when MAE is the optimization goal and NLPD~\citep{laparra2016perceptual} is the evaluation measure for the deblurring task. Nevertheless, MAE ranks its own results the second place. As shown in Sec.~\ref{sec:visual}, the resulting images from MAE and NLPD look visually similar. 
  
  \begin{figure*}
    \centering
      \includegraphics[height=0.25\linewidth]{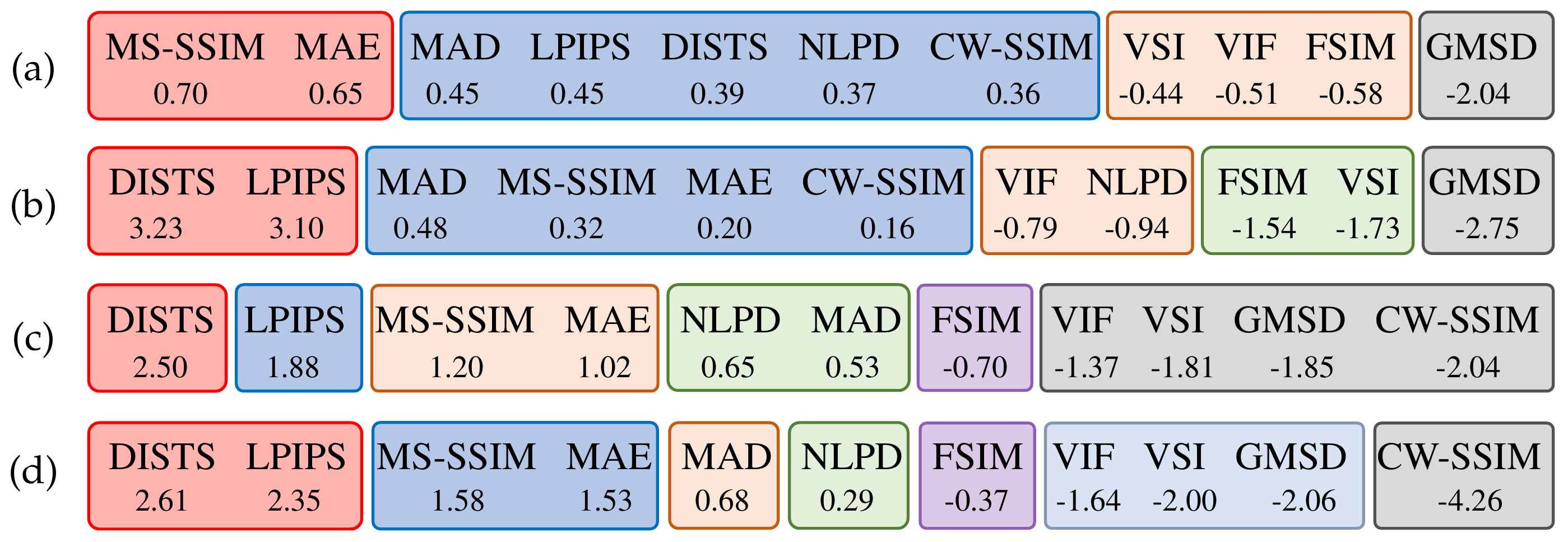}
    \caption{Subjective ranking of the final results in the four tasks, based on human opinion scores. (a) Denoising. (b) Deblurring. (c) Super-resolution. (d) Compression. The optimization performance of IQA models is ranked in the descending order from left to right. Below each model is the global ranking score (larger is better). Models within the same colored box have statistically indistinguishable performance.}
    \label{fig:bt-rank}
  \end{figure*}
  
  \subsection{Subjective Testing} \label{subsec:subjective}
  We conducted an experiment to acquire human perceptual comparisons of the IQA optimized results. A two-alternative forced choice (2AFC) method was employed, allowing differentiation of fine-grained quality variations. On each trial, subjects were shown two images optimized according to two different IQA methods, presented on the left and right side of the corresponding reference image (see Fig. \ref{fig:gui}). Subjects were asked to choose which of the two images had better quality. Subjects were allowed unlimited viewing time, and were free to adjust their viewing distance.  A customized graphical user interface (GUI) was used to display the images at resolution matched to the screen (\ie, $512\times512$ pixels), and subjects were able to zoom in to any portion of the images for more careful comparison.
  The screen had the resolution of $1,920\times 1,080$ pixels, and was calibrated in accordance with the recommendations of ITU-R BT.500-11 \cite{bt2002methodology}. Tests were performed in indoor spaces with ordinary illumination levels. 
  
  We generated a total of $\binom{11}{2}\times 4\times20=4400$ paired comparisons for $11$ IQA models, $4$ tasks, and $20$ test images.  We gathered data from $25$ subjects ($13$ males and $12$ females) aged between $18$ and $22$, with normal or corrected-to-normal visual acuity. Subjects had general background knowledge of image processing and computer vision, but were otherwise na\"{i}ve to the purpose of this study. To reduce fatigue, we performed the experiment in multiple sessions, each consisting of $500$ randomly selected comparisons, with the randomized left-right presentation, and allowed subjects to take a break at any time during the session. Subjects were encouraged, but not required, to participate in multiple sessions. In order to detect subjects that were not properly performing the task, we included $5$ pairs where one image was of unambiguously better quality (\eg, the original and a noisy image). Our intention was to discard the results of subjects who failed in more than one of these pairs, but the results of all subjects turned out to be valid. In total, each image pair was evaluated by at least $5$ subjects, and each IQA model was ranked over $1,000$ times for each vision task. 
  
  \begin{figure*}[t]
    \centering
      \subfloat{\includegraphics[height=0.135\linewidth]{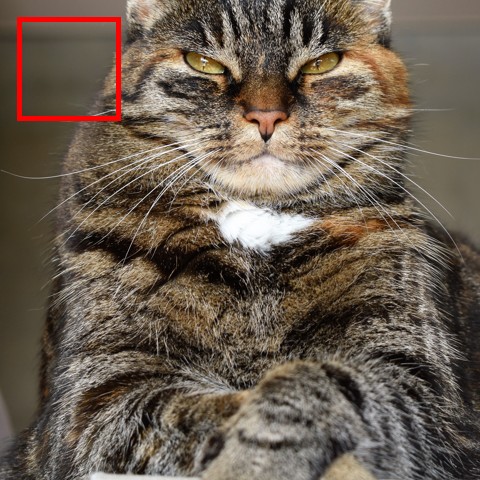}} \hskip.3em
      \subfloat{\includegraphics[height=0.135\linewidth]{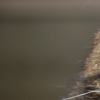}} \hskip.3em
      \subfloat{\includegraphics[height=0.135\linewidth]{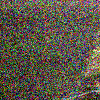}} \hskip.3em
      \subfloat{\includegraphics[height=0.135\linewidth]{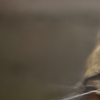}} \hskip.3em
      \subfloat{\includegraphics[height=0.135\linewidth]{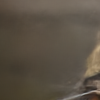}} \hskip.3em
      \subfloat{\includegraphics[height=0.135\linewidth]{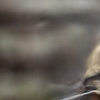}} \hskip.3em
      \subfloat{\includegraphics[height=0.135\linewidth]{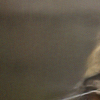}} \\ 
      \addtocounter{subfigure}{-7} \vspace{-1em}
      \subfloat[Original]{\includegraphics[height=0.135\linewidth]{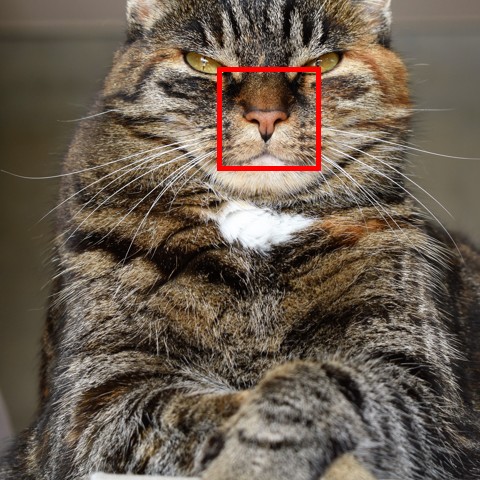}} \hskip.3em
      \subfloat[Cropped]{\includegraphics[height=0.135\linewidth]{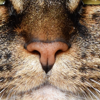}} \hskip.3em
      \subfloat[Noisy]{\includegraphics[height=0.135\linewidth]{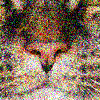}} \hskip.3em
      \subfloat[MAE]{\includegraphics[height=0.135\linewidth]{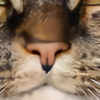}} \hskip.3em
      \subfloat[MS-SSIM]{\includegraphics[height=0.135\linewidth]{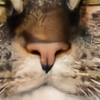}} \hskip.3em
      \subfloat[VIF]{\includegraphics[height=0.135\linewidth]{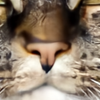}} \hskip.3em
      \subfloat[CW-SSIM]{\includegraphics[height=0.135\linewidth]{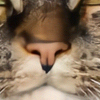}} \\ \vspace{-1em}
      \subfloat{\includegraphics[height=0.135\linewidth]{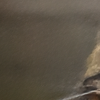}} \hskip.3em
      \subfloat{\includegraphics[height=0.135\linewidth]{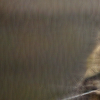}} \hskip.3em
      \subfloat{\includegraphics[height=0.135\linewidth]{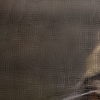}} \hskip.3em
      \subfloat{\includegraphics[height=0.135\linewidth]{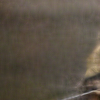}} \hskip.3em
      \subfloat{\includegraphics[height=0.135\linewidth]{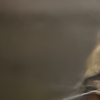}} \hskip.3em
      \subfloat{\includegraphics[height=0.135\linewidth]{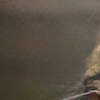}} \hskip.3em
      \subfloat{\includegraphics[height=0.135\linewidth]{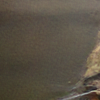}} \\ 
      \addtocounter{subfigure}{-7} \vspace{-1em}
      \subfloat[MAD]{\includegraphics[height=0.135\linewidth]{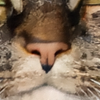}} \hskip.3em
      \subfloat[FSIM]{\includegraphics[height=0.135\linewidth]{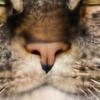}} \hskip.3em
      \subfloat[GMSD]{\includegraphics[height=0.135\linewidth]{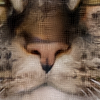}} \hskip.3em
      \subfloat[VSI]{\includegraphics[height=0.135\linewidth]{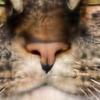}} \hskip.3em
      \subfloat[NLPD]{\includegraphics[height=0.135\linewidth]{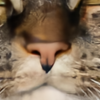}} \hskip.3em
      \subfloat[LPIPS]{\includegraphics[height=0.135\linewidth]{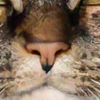}} \hskip.3em
      \subfloat[DISTS]{\includegraphics[height=0.135\linewidth]{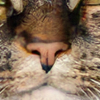}}
    \caption{Denoising results on two regions cropped from an example image, using a DNN optimized for different IQA models (as indicated).} 
    \label{fig:N}
  \end{figure*}

  \section{Experimental Results}\label{sec:results}
  Based on the subjective data, we conducted a quantitative comparison of the IQA models through the lens of perceptual optimization. We also qualitatively compared the visual results associated with the IQA models. Last, we combined a top-performing IQA model with  adversarial loss~\citep{goodfellow2014generative} to test whether additional perceptual gains could be obtained in blind image deblurring.
  
  \subsection{Quantitative Results} \label{subsec:rank_group}
  We employed the Bradley–Terry model~\citep{bradley1952rank} to convert
  paired comparison results to global rankings. This probabilistic model assumes that the visual quality of the $k$-th test image optimized for the $i$-th IQA model, $q^{k}_i$, follows a Gumbel distribution with location $\mu^{k}_i$ and scale $s$. Assuming independence between $q^k_i$ and $q^k_j$, the difference $q^k_i - q^k_j$ is a logistic random variable, and therefore $p^k_{ij} = P(q^k_i \ge q^k_j)$ can be computed using the logistic cumulative distribution function:
  \begin{align}
  p^k_{ij}=P(q^k_i - q^k_j\ge 0) =\frac{\exp(\mu^k_{i}/s)}{\exp(\mu^k_{i}/s)+\exp(\mu^k_{j}/s)},
  \end{align}
  where $s$ is usually set to $1$, leading to a simplified expression:
  \begin{align}
      p^k_{ij} = \frac{e^{\mu^k_i}}{e^{\mu^k_i}+e^{\mu^k_j}}.
  \end{align}
  As such, we may obtain the negative log-likelihood of our pairwise count matrix $W^k$: 
  \begin{align}
  \ell(\mu^k|W^k)=\sum_{i=1}^{M} \sum_{j=1 \atop j \neq i}^{M}\Big( w^k_{i j} \log \left(e^{\mu^k_{i}}+e^{\mu^k_{j}}\right)- w^k_{ij} \mu^k_{i}\Big),\label{eq:nll}
  \end{align}
  where $w^k_{ij}$ represents the number of times that $D_i$ is preferred over $D_j$ for the $k$-th test image. For each of the four low-level vision tasks, we minimized Eq.~\eqref{eq:nll} iteratively using gradient descent to obtain the optimal estimate $\hat{\mu}^k$. We averaged $\hat{\mu}^k$ over the $20$ test images, resulting in four global rankings of perceptual optimization performance, as shown in Fig.~\ref{fig:bt-rank}. It is clear that MS-SSIM~\citep{wang2003multiscale} and MAE are superior to the other IQA models in the task of denoising, whereas DNN-based measures DISTS~\citep{ding2020dists} and LPIPS~\citep{zhang2018unreasonable}, outperform the others in all other tasks. Thus, there is no single IQA model that performs best across all tasks. We ascribe this to differences in the nature of the tasks: denoising requires distinguishing signal and noise,  deblurring, super-resolution, and compression all require recovery of discarded information from partial deterministic measurements (for the first two, via linear projection, and for compression via quantization).  MS-SSIM and MAE are both known to prefer smooth appearances, and are seen to excel at denoising. Both DISTS and LPIPS explicitly represent aspects of fine textures, and are superior for the remaining three tasks.  Finally, it is important to note that many of the models, despite their impressive abilities to explain existing IQA databases, are outperformed by MAE, the simplest metric in our set.
  
  \begin{figure*}[t]
    \centering
      \subfloat{\includegraphics[height=0.135\linewidth]{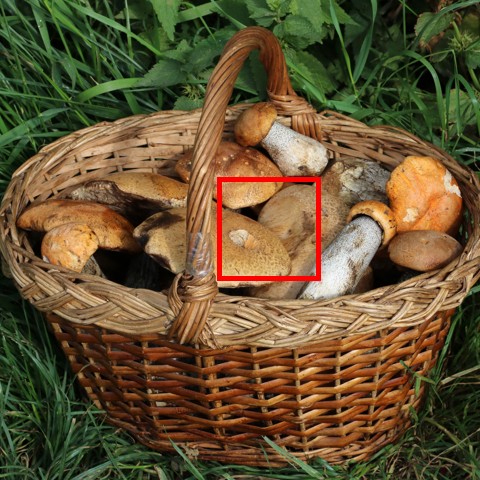}} \hskip.3em
      \subfloat{\includegraphics[height=0.135\linewidth]{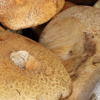}} \hskip.3em
      \subfloat{\includegraphics[height=0.135\linewidth]{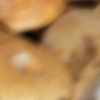}} \hskip.3em
      \subfloat{\includegraphics[height=0.135\linewidth]{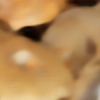}} \hskip.3em
      \subfloat{\includegraphics[height=0.135\linewidth]{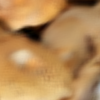}} \hskip.3em
      \subfloat{\includegraphics[height=0.135\linewidth]{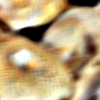}} \hskip.3em
      \subfloat{\includegraphics[height=0.135\linewidth]{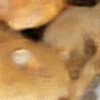}} \\ 
      \addtocounter{subfigure}{-7} \vspace{-1em}
      \subfloat[Original]{\includegraphics[height=0.135\linewidth]{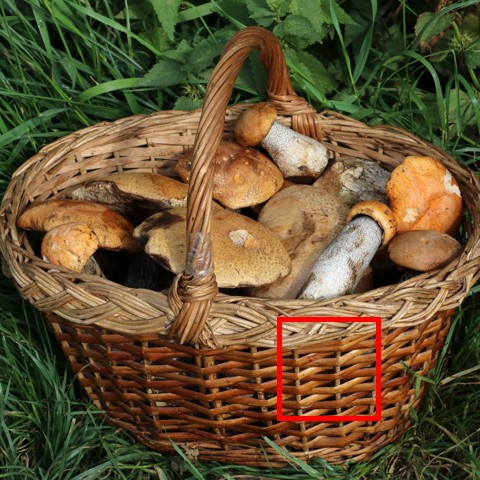}} \hskip.3em
      \subfloat[Cropped]{\includegraphics[height=0.135\linewidth]{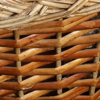}} \hskip.3em
      \subfloat[Blurred]{\includegraphics[height=0.135\linewidth]{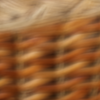}} \hskip.3em
      \subfloat[MAE]{\includegraphics[height=0.135\linewidth]{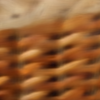}} \hskip.3em
      \subfloat[MS-SSIM]{\includegraphics[height=0.135\linewidth]{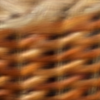}} \hskip.3em
      \subfloat[VIF]{\includegraphics[height=0.135\linewidth]{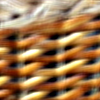}} \hskip.3em
      \subfloat[CW-SSIM]{\includegraphics[height=0.135\linewidth]{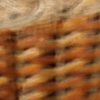}} 
      \\ \vspace{-1em}
      \subfloat{\includegraphics[height=0.135\linewidth]{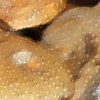}} \hskip.3em
      \subfloat{\includegraphics[height=0.135\linewidth]{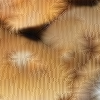}} \hskip.3em
      \subfloat{\includegraphics[height=0.135\linewidth]{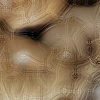}} \hskip.3em
      \subfloat{\includegraphics[height=0.135\linewidth]{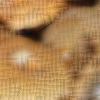}} \hskip.3em
      \subfloat{\includegraphics[height=0.135\linewidth]{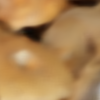}} \hskip.3em
      \subfloat{\includegraphics[height=0.135\linewidth]{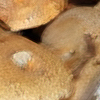}} \hskip.3em
      \subfloat{\includegraphics[height=0.135\linewidth]{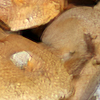}} \\ 
      \addtocounter{subfigure}{-7} \vspace{-1em}
      \subfloat[MAD]{\includegraphics[height=0.135\linewidth]{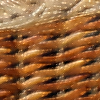}} \hskip.3em
      \subfloat[FSIM]{\includegraphics[height=0.135\linewidth]{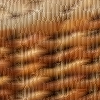}} \hskip.3em
      \subfloat[GMSD]{\includegraphics[height=0.135\linewidth]{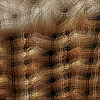}} \hskip.3em
      \subfloat[VSI]{\includegraphics[height=0.135\linewidth]{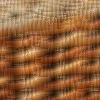}} \hskip.3em
      \subfloat[NLPD]{\includegraphics[height=0.135\linewidth]{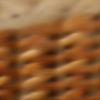}} \hskip.3em
      \subfloat[LPIPS]{\includegraphics[height=0.135\linewidth]{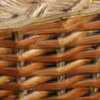}} \hskip.3em
      \subfloat[DISTS]{\includegraphics[height=0.135\linewidth]{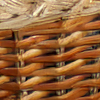}}
      
    \caption{Deblurring results on two regions cropped from an example image, using a DNN optimized for different IQA models.} 
    \label{fig:B}
  \end{figure*}
  
  To determine whether the optimization results of the IQA models are statistically significant, we conducted an independent paired-sample $t$-test. The null hypothesis is that the ranking scores $\{\mu^k_i\}_{k=1}^{20}$ for $D_i$ and $\{\mu^k_j\}_{k=1}^{20}$ for $D_j$ come from the same normal distribution with unknown variance. When the test cannot reject the null hypothesis at the $\alpha=5\%$ significance level, the two IQA models have statistically indistinguishable performance, and we considered them to belong to the same group. Grouping results are shown in Fig.~\ref{fig:bt-rank}. Surprisingly, we find that the perceptual gains of MS-SSIM over MAE are statistically insignificant on all four tasks, despite the fact that MS-SSIM is far better than MAE in explaining existing IQA databases. Relying on similar sets of VGG features~\citep{Simonyan14c}, DISTS and LPIPS also achieve similar performance, except for the super-resolution task where the former is statistically better.
  
  \begin{table}[t]
  \newcommand{\tabincell}[2]{\begin{tabular}{@{}#1@{}}#2\end{tabular}}
    \centering
    \caption{SRCC of objective ranking scores from the IQA models against subjective ranking scores}
      \begin{tabular}{lcccc}
      \toprule
      IQA Model &  \tabincell{c}{Denois-\\ing}  &\tabincell{c}{Deblur-\\ring}   & \tabincell{c}{Super-\\resolution} & \tabincell{c}{Compress-\\ion}\\ \hline 
      MAE     &  0.527 & 0.164 &  0.309 &  0.455\\
      MS-SSIM &  \textbf{0.564} & 0.127 &  0.455 &  0.346\\
      VIF      & 0.273  & 0.600 &  0.418 &  0.018\\
      CW-SSIM   &  0.382 & 0.418 &  0.091 &  0.018\\
      MAD      &  0.418 & 0.455 &  0.346 &  0.382\\
      FSIM      &  0.236 & 0.054 &  0.091 &  0.127\\
      GMSD     &  0.091 & 0.018 &  0.127 &  0.127\\
      VSI      &  0.164 &  0.018 &  0.018 &  0.091\\
      NLPD      &  0.491 & 0.127 &  0.200 &  0.309\\
      LPIPS     &  \textbf{0.709} & \textbf{0.855} &  \textbf{0.782} &  \textbf{0.782}\\
      DISTS     &  0.346 & \textbf{0.891} & \textbf{0.782} &  \textbf{0.855}\\
      \bottomrule
      \end{tabular}
    \label{tab:method_srcc}
  \end{table}
  
  By computing the Spearman's rank correlation coefficient (SRCC) between objective model rankings (in Fig.~\ref{fig:matrix}) and subjective human rankings (in Fig.~\ref{fig:bt-rank}), we are able to compare the algorithm-level performance of the $11$ IQA models on the new dataset. We find from the Table~\ref{tab:method_srcc} that there is a lack of correlation between  model predictions and human judgments for the majority of IQA methods. DISTS and LPIPS tend to rank the images with complex model-dependent distortions in a more perceptually consistent way. We refer interested readers to Appendix~\ref{ap:assess} for 
  more comparisons on several IQA databases dedicated to low-level vision problems.

  \begin{figure*}[t]
    \centering
      \subfloat{\includegraphics[height=0.135\linewidth]{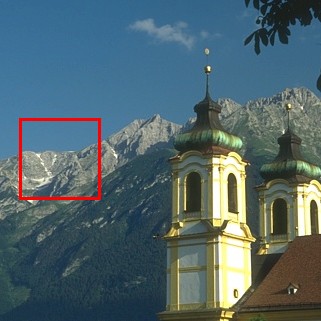}} \hskip.3em
      \subfloat{\includegraphics[height=0.135\linewidth]{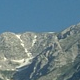}} \hskip.3em
      \subfloat{\includegraphics[height=0.135\linewidth]{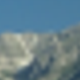}} \hskip.3em
      \subfloat{\includegraphics[height=0.135\linewidth]{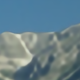}} \hskip.3em
      \subfloat{\includegraphics[height=0.135\linewidth]{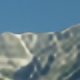}} \hskip.3em
      \subfloat{\includegraphics[height=0.135\linewidth]{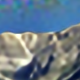}} \hskip.3em
      \subfloat{\includegraphics[height=0.135\linewidth]{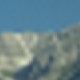}} \\ 
      \addtocounter{subfigure}{-7} \vspace{-1em}
      \subfloat[Original]{\includegraphics[height=0.135\linewidth]{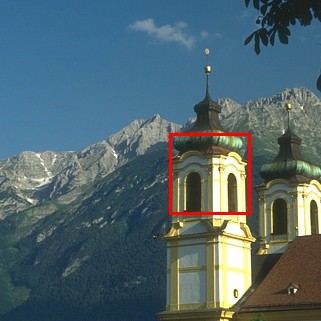}} \hskip.3em
      \subfloat[Cropped]{\includegraphics[height=0.135\linewidth]{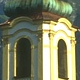}} \hskip.3em
      \subfloat[Low-res]{\includegraphics[height=0.135\linewidth]{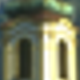}} \hskip.3em
      \subfloat[MAE]{\includegraphics[height=0.135\linewidth]{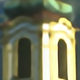}} \hskip.3em
      \subfloat[MS-SSIM]{\includegraphics[height=0.135\linewidth]{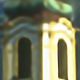}} \hskip.3em
      \subfloat[VIF]{\includegraphics[height=0.135\linewidth]{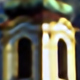}} \hskip.3em
      \subfloat[CW-SSIM]{\includegraphics[height=0.135\linewidth]{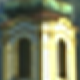}} 
      \\ \vspace{-1em}
      \subfloat{\includegraphics[height=0.135\linewidth]{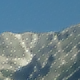}} \hskip.3em
      \subfloat{\includegraphics[height=0.135\linewidth]{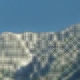}} \hskip.3em
      \subfloat{\includegraphics[height=0.135\linewidth]{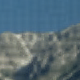}} \hskip.3em
      \subfloat{\includegraphics[height=0.135\linewidth]{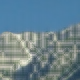}} \hskip.3em
      \subfloat{\includegraphics[height=0.135\linewidth]{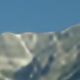}} \hskip.3em
      \subfloat{\includegraphics[height=0.135\linewidth]{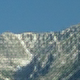}} \hskip.3em
      \subfloat{\includegraphics[height=0.135\linewidth]{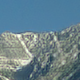}} \\ 
      \addtocounter{subfigure}{-7} \vspace{-1em}
      \subfloat[MAD]{\includegraphics[height=0.135\linewidth]{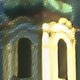}} \hskip.3em
      \subfloat[FSIM]{\includegraphics[height=0.135\linewidth]{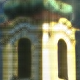}} \hskip.3em
      \subfloat[GMSD]{\includegraphics[height=0.135\linewidth]{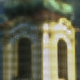}} \hskip.3em
      \subfloat[VSI]{\includegraphics[height=0.135\linewidth]{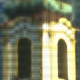}} \hskip.3em
      \subfloat[NLPD]{\includegraphics[height=0.135\linewidth]{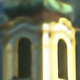}} \hskip.3em
      \subfloat[LPIPS]{\includegraphics[height=0.135\linewidth]{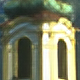}} \hskip.3em
      \subfloat[DISTS]{\includegraphics[height=0.135\linewidth]{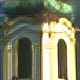}}
    \caption{Super-resolution results on two cropped regions from an example image, using a DNN optimized for different IQA models.}
    \label{fig:S}
  \end{figure*}

  \begin{figure*}[t]
    \centering
      \subfloat{\includegraphics[height=0.135\linewidth]{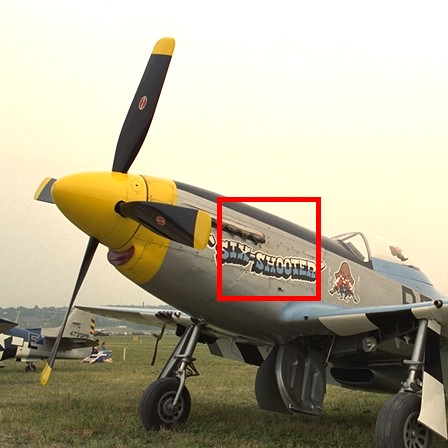}} \hskip.3em
      \subfloat{\includegraphics[height=0.135\linewidth]{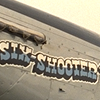}} \hskip.3em
      \subfloat{\includegraphics[height=0.135\linewidth]{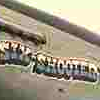}} \hskip.3em
      \subfloat{\includegraphics[height=0.135\linewidth]{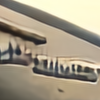}} \hskip.3em
      \subfloat{\includegraphics[height=0.135\linewidth]{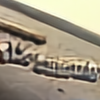}} \hskip.3em
      \subfloat{\includegraphics[height=0.135\linewidth]{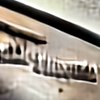}} \hskip.3em
      \subfloat{\includegraphics[height=0.135\linewidth]{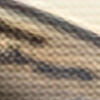}} \\ 
      \addtocounter{subfigure}{-7} \vspace{-1em}
      \subfloat[Original]{\includegraphics[height=0.135\linewidth]{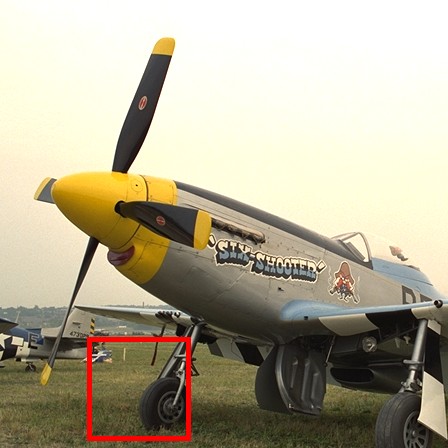}} \hskip.3em
      \subfloat[\scriptsize{Uncompressed}]{\includegraphics[height=0.135\linewidth]{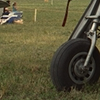}} \hskip.3em
      \subfloat[JPEG]{\includegraphics[height=0.135\linewidth]{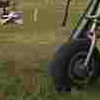}} \hskip.3em
      \subfloat[MAE]{\includegraphics[height=0.135\linewidth]{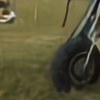}} \hskip.3em
      \subfloat[MS-SSIM]{\includegraphics[height=0.135\linewidth]{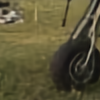}} \hskip.3em
      \subfloat[VIF]{\includegraphics[height=0.135\linewidth]{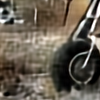}} \hskip.3em
      \subfloat[CW-SSIM]{\includegraphics[height=0.135\linewidth]{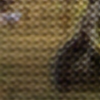}} 
      \\ \vspace{-1em}
      \subfloat{\includegraphics[height=0.135\linewidth]{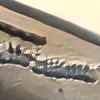}} \hskip.3em
      \subfloat{\includegraphics[height=0.135\linewidth]{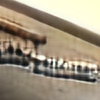}} \hskip.3em
      \subfloat{\includegraphics[height=0.135\linewidth]{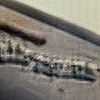}} \hskip.3em
      \subfloat{\includegraphics[height=0.135\linewidth]{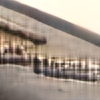}} \hskip.3em
      \subfloat{\includegraphics[height=0.135\linewidth]{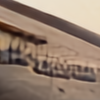}} \hskip.3em
      \subfloat{\includegraphics[height=0.135\linewidth]{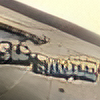}} \hskip.3em
      \subfloat{\includegraphics[height=0.135\linewidth]{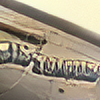}} \\ 
      \addtocounter{subfigure}{-7} \vspace{-1em}
      \subfloat[MAD]{\includegraphics[height=0.135\linewidth]{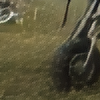}} \hskip.3em
      \subfloat[FSIM]{\includegraphics[height=0.135\linewidth]{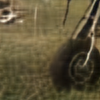}} \hskip.3em
      \subfloat[GMSD]{\includegraphics[height=0.135\linewidth]{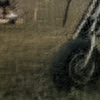}} \hskip.3em
      \subfloat[VSI]{\includegraphics[height=0.135\linewidth]{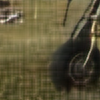}} \hskip.3em
      \subfloat[NLPD]{\includegraphics[height=0.135\linewidth]{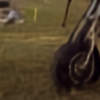}} \hskip.3em
      \subfloat[LPIPS]{\includegraphics[height=0.135\linewidth]{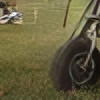}} \hskip.3em
      \subfloat[DISTS]{\includegraphics[height=0.135\linewidth]{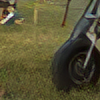}}
    \caption{Compression results on two cropped regions from an example image, using a DNN optimized for different IQA models.} 
    \label{fig:C}
  \end{figure*}
  
  \subsection{Qualitative Results} \label{sec:visual}
  In this subsection, we show example images produced by each IQA-optimized method, qualitatively summarize the types of visual distortion, and use them to diagnose the shortcomings of the corresponding IQA models.
  More visual examples can be found in Figs. \ref{fig:ap_n} - \ref{fig:ap_c}.
  
  Fig.~\ref{fig:N} shows \textit{denoising} results for the ``cat'' image. We observe that MAE, MS-SSIM, and NLPD do a good job in denoising flat regions, but tend to over-smooth texture regions. VIF encourages detail enhancement, leading to artificial local contrast, while GMSD produces a relatively dark appearance presumably because it discards local luminance information. Moreover,  the results of FSIM and VSI exhibit  noticeable artifacts. LPIPS and DISTS preserve fine details, but may not fully remove noise in smooth regions, mistaking the remaining noise as visually plausible texture. Overall, traditional IQA models MAE and MS-SSIM denoise images with various content variations robustly, keeping high-frequency information loss within the acceptable range. This may explain why they are the dominant objective functions for this task.
  
  Fig.~\ref{fig:B} shows \textit{deblurring} results for the ``basket'' image. We see that most of the IQA methods fail, but in different ways. Specifically, the results of MAE, MS-SSIM, CW-SSIM, and NLPD are  quite blurred. FSIM, GMSD, and VSI generate severe ringing artifacts. VIF again fails to adjust the local contrast. MAD exhibits undesirable white dot artifacts, although the main structures are sharp. LPIPS succeeds in deblurring this example, while DISTS produces a result that is closest to the original. This is consistent with current state-of-the-art deblurring results \citep{kupyn2019deblurgan}, generated by incorporating comparison of  the VGG features into the loss.
  
  Fig.~\ref{fig:S} shows \textit{super-resolution} results for the  ``corner tower'' image. Again, MAE, MS-SSIM, NLPD, and especially CW-SSIM produce somewhat blurred images, without recovering fine details. MAD, FSIM, GMSD, and VSI are able to generate some ``structures'', but these are perceived as unpleasant model-dependent artifacts. Benefiting from its texture synthesis capability, DISTS has the potential to super-resolve perceptually plausible fine details, although they differ from those of the original image. 
  
  Fig.~\ref{fig:C} shows \textit{compression} results for the ``airplane'' image at  $0.24\pm 0.01$  bpp. A JPEG image, compressed to $0.25$ bpp, suffers from block and blur artifacts. Overall, the main structures of the original image are well preserved for most IQA models, but the fine details (\eg, the grass) have to be discarded at this low bitrate, or are synthesized with other forms of distortion. VIF reconstitutes a desaturated image with over-enhanced global contrast, and CW-SSIM superimposes periodic artifacts on the underlying image. White dots and ringing artifacts are again apparent in the results of MAD and VSI, respectively. The image by NLPD is blurred and red-shifted. Both LPIPS and DISTS succeed in synthesizing textures that are visually similar to the original.

  \begin{figure*}
    \centering
      \subfloat{\includegraphics[height=0.36\linewidth]{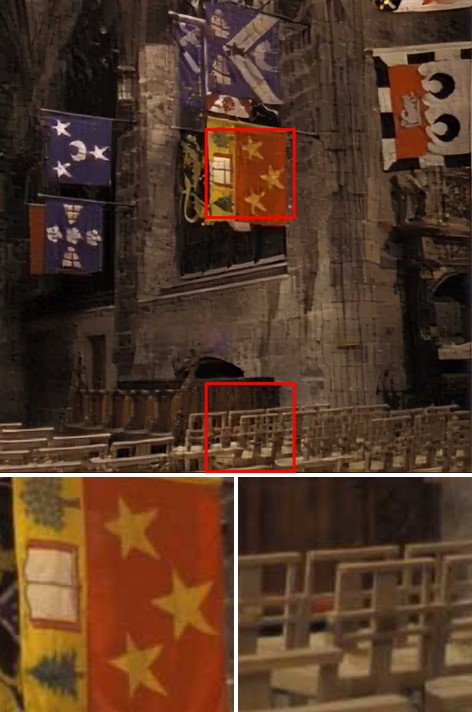}} \hskip.4em
      \subfloat{\includegraphics[height=0.36\linewidth]{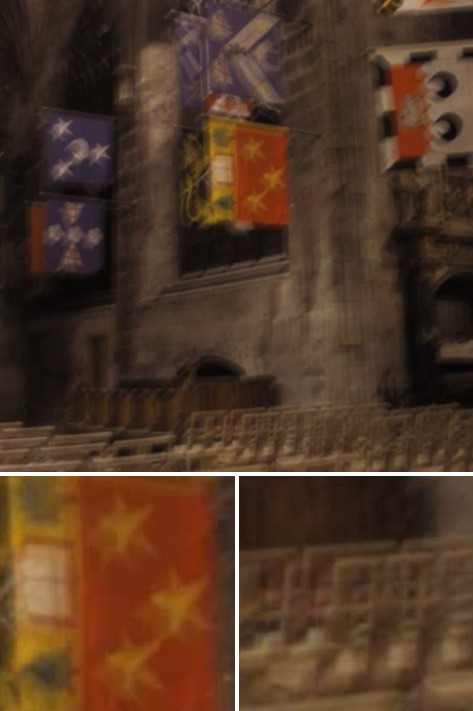}} \hskip.4em
      \subfloat{\includegraphics[height=0.36\linewidth]{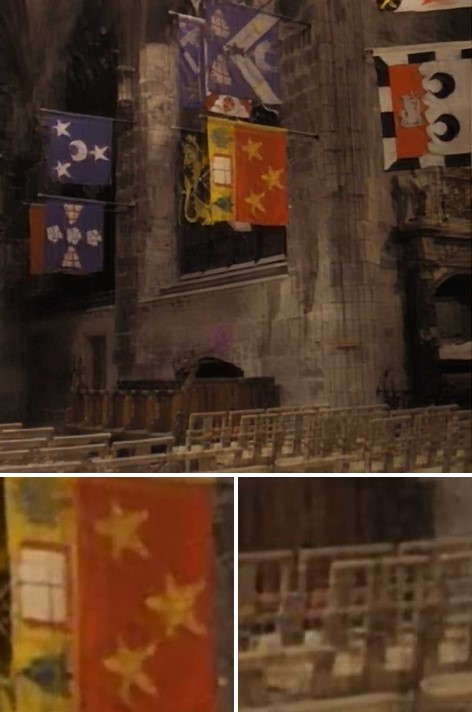}} \hskip.4em
      \subfloat{\includegraphics[height=0.36\linewidth]{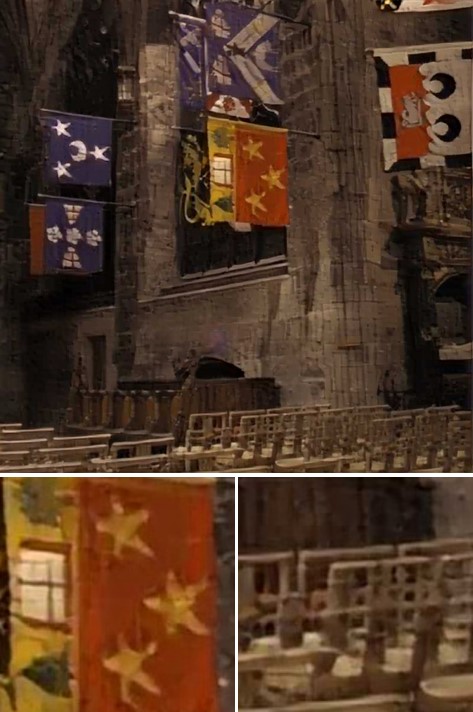}} \\ \vspace{-0.8em}
      \addtocounter{subfigure}{-4} 
      \subfloat[Clean]{\includegraphics[height=0.36\linewidth]{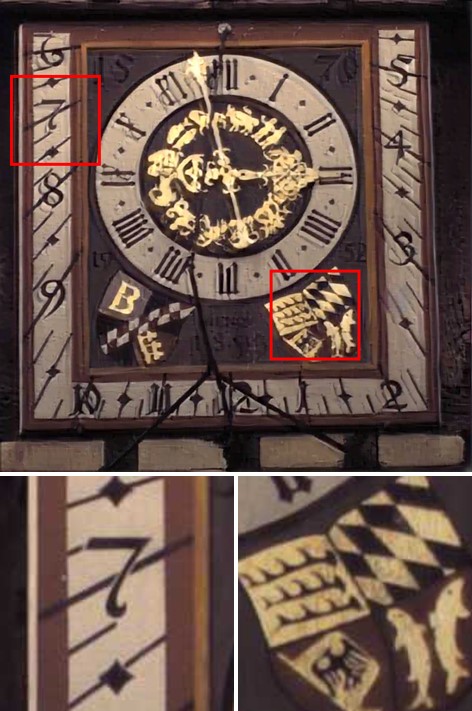}} \hskip.4em
      \subfloat[Blurry]{\includegraphics[height=0.36\linewidth]{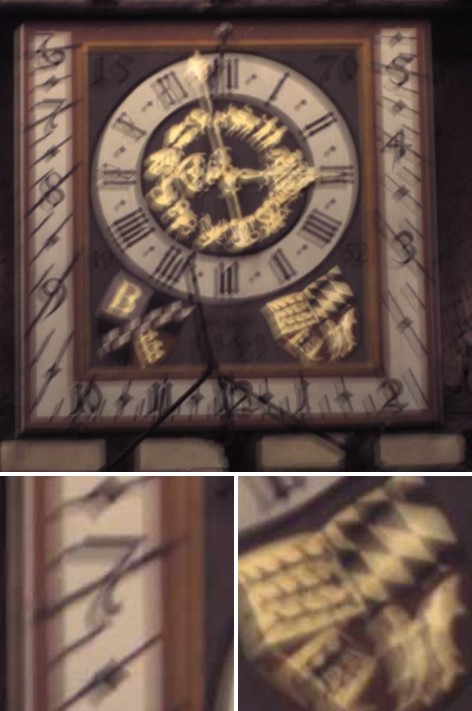}} \hskip.4em
      \subfloat[DeblurGAN-v2]{\includegraphics[height=0.36\linewidth]{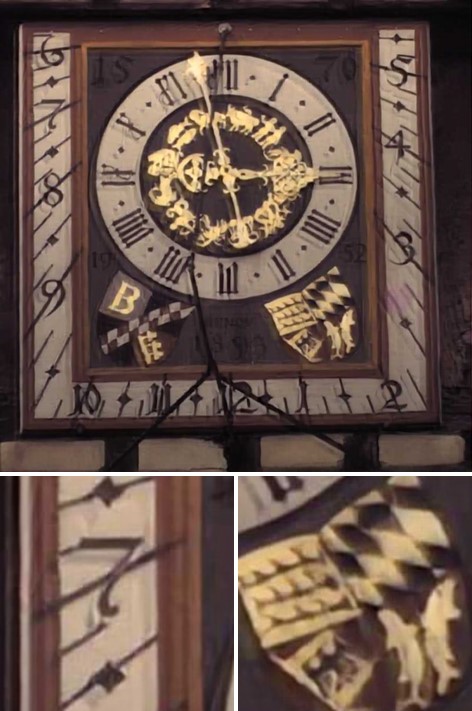}} \hskip.4em
      \subfloat[Fine-tuned]{\includegraphics[height=0.36\linewidth]{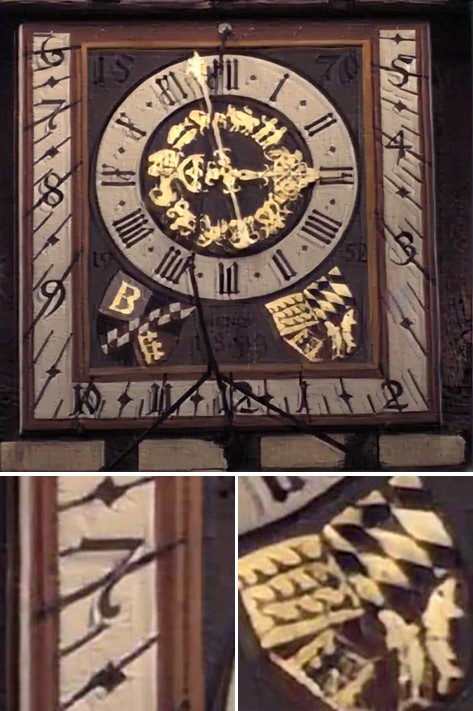}} 
    \caption{Deblurring examples obtained by the original DebluGAN-v2 and the fine-tuned DeblurGAN-v2 (with the loss in Eq.~\eqref{eq:ln}).} 
    \label{fig:deblurgan}
  \end{figure*}
  
  We can summarize the artifacts created during perceptual optimization, some of which are not found in traditional image databases for the purpose of quality assessment:
  
  \begin{itemize}
  \item \textit{Blurring} is a frequently seen distortion type in all four of the tasks, and is mainly caused by error visibility methods (\eg, MAE and NLPD) and structural similarity methods (\eg, MS-SSIM), which rely on simple injective mappings. Specifically, MAE and SSIM work directly with pixels, and NLPD transforms the input image to a multi-scale overcomplete representation using a single stage of local mean subtraction and divisive normalization.
  Under strict constraints imposed by the tasks, they prefer to make a more conservative estimate, producing something akin to an average of all possible outcomes with sharp structures, as would occur when optimizing MSE.
  
  \item \textit{Ringing} is a high-frequency distortion type that often occurs in the images optimized for FSIM, VSI and GMSD (see Fig.~\ref{fig:B} (i) - (k)).  One common characteristic of the three models is that they rely heavily (in some cases, solely) on local gradient magnitude for feature similarity comparison, underweighting (or abandoning) other perceptually important features (such as local luminance and local phase). This creates  ``shortcuts'' that the DNNs can exploit, generating distortions with similar local gradient statistics.

  \item \textit{White dot} artifacts are typical in the optimization results of MAD, which extracts lower-order image statistics from responses of Gabor filters at multiple scales and orientations. The resulting set of statistical measurements seems insufficient to summarize natural image structures that exhibit higher-order dependencies. Therefore, MAD is ``blind'' to distortions that satisfy the same set of statistical constraints, and gives the optimized distorted image a high-quality score.  
  
  \item \textit{Over-enhancement} of local image contrast is encouraged by VIF, which, in most of our experiments, causes significant quality degradation. We believe this arises because VIF does not fully respect reference information when normalizing the covariance term. Specifically, only the second-order statistics of the reference image are used to construct the normalization factor. By incorporating the same statistics computed from the distorted image into normalization, the problem of over-enhancement may be alleviated. In general, quality assessment of image enhancement is a challenging problem~\citep{fang2015no,wang2015patch}, and to the best of our knowledge, all existing full-reference IQA models fail to reward properly-enhanced cases, while penalizing over-enhanced cases.
  
  \item \textit{Luminance and color} artifacts are perceived in final images that are associated with many IQA models. Two causes seem plausible. First, methods such as GMSD discard luminance information. Second, methods such as MS-SSIM and NLPD are originally designed for grayscale images only. Applying them to RGB channels separately fails to take into account hue and saturation information. Transforming to a perceptually better color space, and making use of knowledge of color distortions~\citep{rajashekar2009quantifying} offers an opportunity for improvement. 
  
  \end{itemize}
  
  \subsection{Combining with Adversarial Loss} \label{sec:comb}
  In the field of image restoration and generation, many state-of-the-art algorithms are based on adversarial training~\citep{goodfellow2014generative},  demonstrating impressive capabilities in synthesizing realistic visual content. The output of the adversarial loss is the probability of an image being computer-generated, but this does not confer capabilities for  no-reference IQA modeling, as confirmed by a low SRCC of $0.366$ on the LIVE dataset \citep{LIVE}. 
  Nevertheless, adversarial loss may be useful at the algorithm level, meaning that given a set of images generated by a computational method, the average probability quantitatively measures the capability of the method in generating photorealistic high-quality images. In this subsection, we explored the combination of the adversarial loss and a top-performing IQA measure for additional perceptual gains.

  We chose the task of blind image deblurring, and fine-tuned a state-of-the-art model - DeblurGAN-v2 (under the configuration of Inception-ResNet) \citep{kupyn2019deblurgan}. The original loss function for the generator is
  \begin{align}
      \ell_o = 0.5 \times \ell_{\mathrm{MSE}}+0.006\times \ell_{\mathrm{VGG}}+0.01\times \ell_{\mathrm{Adv}}.
      \label{eq:lo}
  \end{align}
  The first and second terms are the MSE on pixels and responses of conv3\_3 of VGG19~\citep{Simonyan14c}, respectively, and  $\ell_\mathrm{Adv}$ is a variant  of the adversarial loss \citep{kupyn2019deblurgan}. 
  We selected the best-performing IQA model - DISTS - for this experiment. We followed the same training strategy, but modified the loss function of the generator to be
  \begin{align}
      \ell_n = \ell_{\mathrm{DISTS}}+0.001\times \ell_{\mathrm{Adv}},
      \label{eq:ln}
  \end{align}
  where $\ell_{\mathrm{DISTS}}$ denotes the DISTS index. An immediate advantage of this replacement is that the number of hyperparameters is reduced, making manual hyperparameter adjustment easier. After fine-tuning, the average DISTS value decreases from $0.22$ to $0.18$ on the K{\"o}hler test dataset \citep{kohler2012recording}.
  Fig.~\ref{fig:deblurgan} shows two visual examples, from which we find that the fine-tuned results have sharper edges and enhanced contrast, indicating that perceptual gains may be obtained by DISTS on the two examples.
  
  
  \section{Conclusions}
  We have conducted a comprehensive study of perceptual optimization of four low-level vision tasks, guided by  eleven full-reference IQA models.  
  This provides an alternative means of testing the perceptual relevance of IQA models in a practical setting, which we believe is an important complement to the conventional methodology for IQA model evaluation. Subjective testing led to several useful findings. First, through perceptual optimization, we generated a number of distortions (different from those used in existing IQA databases), which may easily fool the respective  models  or models of similar design philosophies (see Table \ref{tab:method_srcc}). 
  It should be noted that the emergence of specific distortions is in principle dependent on the experimental choices (\eg, initialization strategy, model architecture, and optimization technique). Second, although they underperformed the DNN-based models on three of four applications, the standard full-reference IQA models (MS-SSIM and MAE) are still valuable tools for optimizing image processing systems due to their robustness and simplicity. Third, more recent IQA models with surjective mappings may still be used to monitor image quality and to optimize the parameter settings of image processing methods, but in a limited and well-controlled space. Last, the two DNN-based models (LPIPS and DISTS) offered the best overall performance in our experiments, but their high computational complexity and lack of interpretability may hinder their use.
  
  Our work has interesting connections to two separate lines of research. 
  First, inspired by the philosophy of ``analysis by synthesis''~\citep{grenander1970unified}, \cite{wang2008maximum} introduced the maximum differentiation competition methodology to automatically synthesize images for efficiently comparing IQA models. Given two IQA models, MAD generates samples in the space of all possible images that best discriminate the two models. However, the synthesized images may be highly unnatural, and in this case, of limited practical importance. \cite{ma2019group} alleviated this issue  by manually constraining the search space to a finite image set of practical interest. Our approach combines the best aspects of these two methods, in the sense that the test images for model comparison are automatically generated by the trained networks, but arise as solutions of real-world vision tasks and are thus of practical importance. Second, the existence of type II adversarial examples~\citep{szegedy2013intriguing} has exposed the vulnerability of many computer vision algorithms, where a tiny change to the input that is imperceptible to the human eye would cause the algorithm to make classification mistakes. In our case, weaknesses in an IQA model are exposed through optimized images that may be interpreted as type I ``adversarial'' examples of the model: a significant change is made to the original image that substantially degrades its perceptual quality, but the model still claims that this image is of high quality.
  
  \begin{table*}
    \centering
    \caption{Verification of results obtained by our PyTorch re-implementations of the tested IQA models, on three IQA databases.  Numbers indicate SRCC values (reported in original publication / produced by our re-implementation). Bold indicates methods that are computed only on grayscale images in their original versions; we have extended them to evaluate RGB images by averaging the values across all channels.}
      \begin{tabular}{lcccccc}
      \toprule
        \multirow{2}{*}[-3pt]{IQA Model} & \multicolumn{3}{c}{Grayscale}&\multicolumn{3}{c}{Color}\\ 
        \cmidrule(lr){2-4} \cmidrule(lr){5-7}
       & LIVE & CSIQ & TID2013 & LIVE & CSIQ & TID2013\\ \hline 
       \textbf{MS-SSIM}  & 0.951 / 0.951 & 0.906 / 0.886 & 0.786 / 0.782 & 0.931 / 0.932 & 0.902 / 0.886 & 0.801 / 0.816 \\
       \textbf{CW-SSIM}  & 0.786 / 0.781 & 0.745 / 0.738 & 0.673 / 0.680 & 0.741 / 0.747 & 0.744 / 0.744 & 0.709 / 0.725 \\
       \textbf{VIF}  & 0.964 / 0.963 & 0.911 / 0.911 & 0.677 / 0.676 & 0.957 / 0.957 & 0.894 / 0.894 & 0.654 / 0.654\\
       \textbf{NLPD} & 0.937 / 0.938 & 0.932 / 0.937 & 0.800 / 0.800 & 0.917 / 0.914 & 0.913 / 0.913 & 0.812 / 0.808\\
       \textbf{GMSD} & 0.960 / 0.960 & 0.950 / 0.950 & 0.804 / 0.804 & 0.949 / 0.948 & 0.937 / 0.934 & 0.830 / 0.823   \\
       \textbf{MAD} & 0.967 / 0.960 & 0.947 / 0.941 & 0.781 / 0.773 & 0.954 / 0.951 & 0.937 / 0.935 & 0.758 / 0.740   \\   
       FSIM & 0.963 / 0.963 & 0.924 / 0.916 & 0.802 / 0.802 & 0.965 / 0.965 & 0.931 / 0.923 & 0.851 / 0.851 \\
       VSI & 0.953 / 0.950 & 0.930 / 0.923 & 0.805 / 0.793 & 0.952 / 0.956 & 0.942 / 0.937 & 0.897 / 0.889\\
       LPIPS & 0.932 / 0.932 & 0.837 / 0.837 & 0.616 / 0.616 & 0.932 / 0.932 & 0.876 / 0.876 & 0.670 / 0.670\\
       DISTS & 0.942 / 0.942 &  0.905 / 0.905 & 0.764 / 0.764 & 0.954 / 0.954 & 0.929 / 0.929 & 0.830 / 0.830\\
      \bottomrule
      \end{tabular}
    \label{tab:re_srcc}
  \end{table*}

  The analysis of our experimental results suggests several desirable properties that should be included in future IQA methods. First, the transformation used in the IQA model should be perceptual, mapping the input images into a space where a simple distance measure (e.g., Euclidean) matches human judgements of image quality. This is in the same spirit that color scientists pursue perceptually uniform color spaces, and is an underlying principle of a number of existing models (e.g., NLPD). \cite{zhang2018unreasonable} and \cite{ding2020dists} demonstrated that a cascade of linear convolution, downsampling, and rectified nonlinearity optimized for high-level vision tasks may be a good candidate. Second, the IQA model should enjoy unique optima (\ie, the underlying mapping should be injective) to guarantee that images close to optimal are visually similar to the original. This criterion was respected by early models (\eg, MS-SSIM), but was largely overlooked in recent IQA model development.
  Third, the IQA model should be continuous and differentiable, with well-behaved gradients, to aid optimization in complex situations (\eg, training DNNs with millions of parameters). Last but not least, the IQA model should be computationally efficient, enabling real-time quality assessment and perceptual optimization.
  To the best of our knowledge, although many current IQA models possess subsets of these properties, no current IQA model satisfies them all. 
  
  
  \begin{table*}
    \centering
    \caption{Summary of datasets for evaluating full-reference IQA models. Traditional distortion types include artificial Gaussian noise, Gaussian blur, JPEG compression, etc. As in the dataset we describe in Section \ref{subsec:details}, BAPPS contains multiple distortion types, produced by computational methods for different vision tasks.}
      \begin{tabular}{lcccc}
      \toprule
       Dataset & \# of reference images & \# of distorted images & Distortion types\\ \hline 
       LIVE~\citep{LIVE}  & 29 & 779 & Traditional \\
       CSIQ~\citep{larson:011006}  & 30 & 866 & Traditional  \\
       TID2013~\citep{Ponomarenko201557} & 25 & 3,000 & Traditional  \\ \hline
       FLT~\citep{egiazarian2018statistical} & 75 & 300 & Denoising \\
       Liu13~\citep{liu2013no} & 40 & 1,200 & Deblurring  \\
       Lai16~\citep{lai2016comparative} & 25 & 2,800 & Deblurring  \\
       Ma17~\citep{ma2017learning} & 30 & 1,620 & Super-resolution \\
       QADS~\citep{zhou2019visual} & 20 & 980 & Super-resolution \\
       SHRQ~\citep{min2019quality} & 80 & 600 & Dehazing \\
       Tian19 \citep{tian2018benchmark} & 10 & 140 & Rendering \\
       SynTex~\citep{golestaneh2015effect} & 21 & 105 & Texture synthesis  \\
       TQD~\citep{ding2020dists} & 10 & 150 & Texture synthesis  \\
       BAPPS~\citep{zhang2018unreasonable} & - & 26,904 & Multiple  \\
       Proposed & 20 & 880 & Multiple \\
      \bottomrule
      \end{tabular}
    \label{tab:database}
  \end{table*}

  \begin{appendices}
  
  \section{Perceptual Correlation Comparison of IQA Models} \label{ap:assess}
  A conventional method for evaluating IQA models is to compute their agreement with subjective scores in one or more standardized IQA databases (\eg, LIVE~\citep{LIVE}, CSIQ~\citep{larson:011006} or TID2013~\citep{Ponomarenko201557}), consisting of artificially distorted images. Many existing IQA models achieve impressive correlation with these databases (see Table~\ref{tab:re_srcc}), but their performance in assessing the perceptual quality of images produced by low-level vision algorithms has not been tested. In this appendix, we tested them on multiple human-rated image generation/restoration databases, including a denoising database - FLT~\citep{egiazarian2018statistical}, two motion deblurring databases - Liu13~\citep{liu2013no} and Lai16~\citep{lai2016comparative}, two super-resolution databases - Ma17~\citep{ma2017learning} and QADS~\citep{zhou2019visual}, a dehazing database - SHRQ~\citep{min2019quality}, a depth image-based rendering database  - Tian19 \citep{tian2018benchmark}, two texture synthesis databases - SynTex~\citep{golestaneh2015effect} and TQD~\citep{ding2020dists}, and a patch similarity database - BAPPS~\citep{zhang2018unreasonable}. The 
  details of these databases are summarized in Table \ref{tab:database}.
  
  \begin{table*}
    \centering
    \caption{SRCC comparison of IQA models on existing image generation/restoration databases.}
      \begin{tabular}{lcccccccccc}
      \toprule
        \multirow{2}{*}[-3pt]{IQA Model} & 
        \multicolumn{1}{c}{Denoising}&
        \multicolumn{2}{c}{Deblurring} &
        \multicolumn{2}{c}{Super-resolution}&
        \multicolumn{1}{c}{Dehazing}&
        \multicolumn{1}{c}{Rendering}&
        \multicolumn{2}{c}{Texture synthesis}\\ 
        \cmidrule(lr){2-2}\cmidrule(lr){3-4}\cmidrule(lr){5-6}
        \cmidrule(lr){7-7}\cmidrule(lr){8-8}\cmidrule(lr){9-10}
        &  FLT & Lai16 & Liu13 & Ma17 & QADS & SHRQ & Tian19 & SynTEX & TQD \\ \hline 
       PSNR   & 0.183 & 0.301 & 0.803 & 0.592 & 0.360 & 0.740 & 0.536 & 0.114 & 0.233\\
       SSIM   & 0.355 & 0.298 & 0.777 & 0.624 & 0.529 & 0.692 & 0.230 &  0.620 & 0.307\\
       MS-SSIM  & 0.246 & 0.320 & 0.898 & 0.795 & 0.717 & 0.687 & 0.396 &  0.469 & 0.288\\
       VIF   & 0.169 & 0.261 & 0.864 & 0.831 & \textbf{0.815} & 0.667 & 0.259 & 0.448 & 0.305\\
       CW-SSIM  & 0.101 & 0.600 & 0.742 & 0.706 & 0.474 & 0.698 & 0.522 &  0.496 & 0.325\\
       MAD   & 0.182 & 0.446 & 0.897 & \textbf{0.864} & 0.723 & 0.605 & \textbf{0.622} & 0.134 & 0.302\\
       FSIM   & 0.555 & 0.297 & \textbf{0.921} & 0.747 & 0.687 & 0.695 &  0.476 & 0.093 & 0.176\\
       GMSD   & 0.389 & 0.174 & 0.918 & 0.851 & 0.765 & 0.663 & 0.479 & 0.006 & 0.256\\
       VSI   & 0.528 & 0.295 & 0.920 & 0.710 & 0.584 & 0.696 & 0.531 & 0.123 & 0.179\\
       NLPD   & 0.151 & 0.323 & 0.853 & 0.732 & 0.591 & 0.608 & 0.463 & 0.483 & 0.271\\
       PieAPP  & \textbf{0.629} & \textbf{0.601} & 0.786& 0.771 & \textbf{0.849} & 0.725  & 0.298 & \textbf{0.709} & \textbf{0.713}\\
       LPIPS  & 0.457 & 0.347 & 0.867  & 0.788 & 0.669 & \textbf{0.777}& 0.311 & 0.663 & 0.392\\
       DISTS  & \textbf{0.636} & \textbf{0.754} & \textbf{0.941} & \textbf{0.878} & 0.809 & \textbf{0.789} & \textbf{0.671} & \textbf{0.923} & \textbf{0.910}\\
      \bottomrule
      \end{tabular}
    \label{tab:srcc_cmp2}
  \end{table*}

  \begin{table*}
  \newcommand{\tabincell}[2]{\begin{tabular}{@{}#1@{}}#2\end{tabular}}
    \centering
    \caption{2AFC score comparison of IQA models on the BAPPS dataset and the proposed dataset.}
      \begin{tabular}{lcccccccc}
      \toprule
       \multirow{2}{*}[-6pt]{IQA Model} & \multicolumn{4}{c}{BAPPS}& \multicolumn{4}{c}{Proposed} \\
       \cmidrule(lr){2-5} \cmidrule(lr){6-9}
       & \tabincell{c}{Color-\\ization}  & \tabincell{c}{Video\\ deblurring} & \tabincell{c}{Frame\\interpolation} & \tabincell{c}{Super-\\resolution} &
       Denoising & Deblurring & \tabincell{c}{Super-\\resolution} & Compression\\ \hline 
       PSNR  & 0.624 & 0.590 & 0.543 & 0.642 & 0.627 & 0.518 & 0.612 & 0.689\\
       SSIM  & 0.522 & 0.583 & 0.548 & 0.613 & \textbf{0.636} & 0.575 & 0.599 & 0.649\\
       MS-SSIM & 0.522 & 0.589 & 0.572 & 0.638 & 0.623 & 0.568 & 0.655 & 0.665\\
       VIF  & 0.515 & 0.594 & 0.597 & 0.651 & 0.589 & 0.607 & 0.655 & 0.540\\
       CW-SSIM & 0.512 & \textbf{0.601} & 0.604 & 0.665&0.623& 0.651 & 0.584 & 0.496\\
       MAD  & 0.490 & 0.593 & 0.581 & 0.655 & 0.624 & 0.671 & 0.681 & 0.651\\
       FSIM  & 0.573 & 0.590 & 0.581 & 0.660 & 0.522 & 0.490 & 0.525 & 0.563\\
       GMSD  & 0.517 & 0.594 & 0.575 & 0.676 & 0.417 & 0.454 & 0.469 & 0.567\\
       VSI  &  0.597 & 0.591 & 0.568 & 0.668 & 0.518 & 0.470 & 0.487 & 0.576 \\
       NLPD  & 0.528 & 0.584 & 0.552 & 0.655 & 0.622 & 0.514 & 0.629 & 0.652\\
       PieAPP & 0.594 &  0.582 & 0.598 & 0.685 & 0.625 & 0.734 &  \textbf{0.744} & 0.822\\
       LPIPS  & \textbf{0.625} & \textbf{0.605} & \textbf{0.630} & \textbf{0.705} & \textbf{0.657} & \textbf{0.788} & \textbf{0.768} & \textbf{0.834} \\
       DISTS & \textbf{0.627} &  0.600 &  \textbf{0.625} & \textbf{0.710} & 0.602 &  \textbf{0.790} &   0.704 &  \textbf{0.833} \\
      \bottomrule
      \end{tabular}
    \label{tab:2afc_bapps_our}
  \end{table*}

  Tables~\ref{tab:srcc_cmp2} and \ref{tab:2afc_bapps_our} show the performance comparisons of $13$ IQA methods in terms of the SRCC and 2AFC scores. 
  As suggested in  \citep{zhang2018unreasonable}, the 2AFC score is computed by: $pq+(1-p)(1-q)$, where $p$ is the percentage of human votes and $q=\{0,1\}$ is the vote of an IQA model. When $q$ agrees with the majority of human votes, the 2AFC score is larger, indicating better performance.
  We find that the overall performance of all models is lower compared to that in the standard IQA databases (see Table~\ref{tab:re_srcc}), indicating the difficulty of generalizing to unseen distortions. Moreover, DNN-based measures are relatively better than knowledge-driven models in these application-oriented databases, but there is still significant room for improvement\ 
  
\begin{figure*}
    \centering
      \subfloat[Original: PSNR$\uparrow$ / SSIM$\uparrow$ \newline LPIPS$\downarrow$ / DISTS$\downarrow$] {\includegraphics[height=0.175\linewidth]{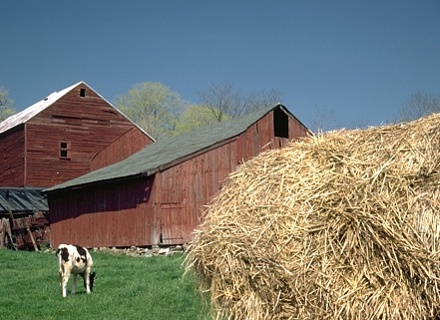}} \hskip.2em
      \subfloat[Bicubic: $20.46$ / $0.577$ \newline $0.373$ / $0.256$]{\includegraphics[height=0.175\linewidth]{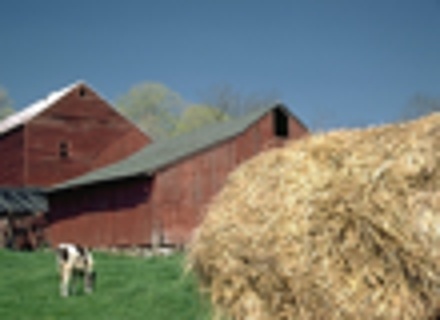}}\hskip.2em
      \subfloat[Glasner09: $19.79$ / $0.602$ \newline $0.400$ / $0.266$]{\includegraphics[height=0.175\linewidth]{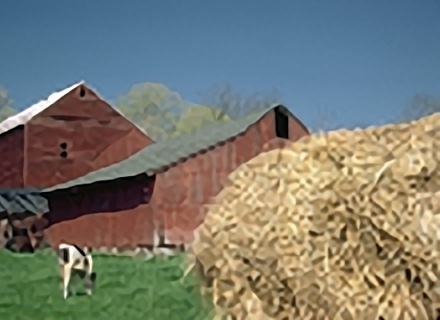}}\hskip.2em
      \subfloat[Yang13: $20.26$ / $0.600$ \newline $0.352$ / $0.232$]{\includegraphics[height=0.175\linewidth]{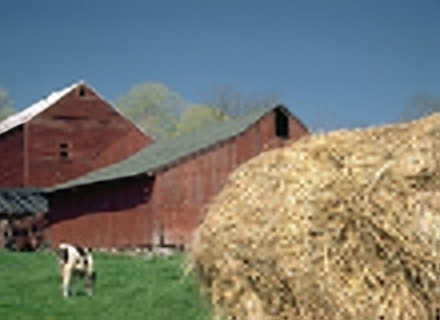}} \\ \vspace{-0.8em} 
      \subfloat[EDSR: $21.10$ / $0.651$ \newline $0.330$ / $0.218$]{\includegraphics[height=0.175\linewidth]{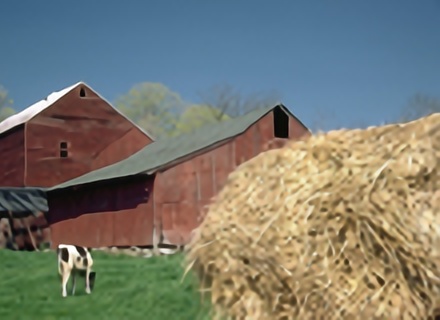}}\hskip.2em
      \subfloat[SRGAN: $17.41$ / $0.546$ \newline $0.357$ / $0.178$]{\includegraphics[height=0.175\linewidth]{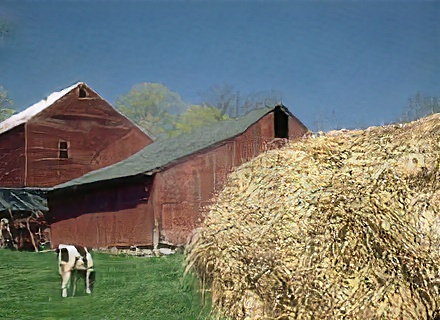}}\hskip.2em
      \subfloat[ESRGAN: $17.63$ / $0.550$ \newline $0.239$ / $0.133$]{\includegraphics[height=0.175\linewidth]{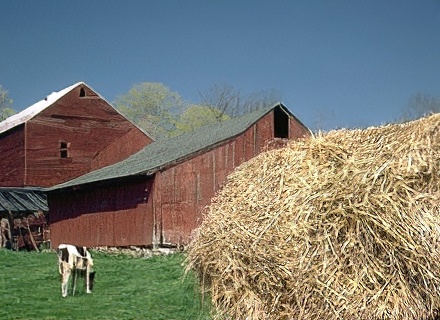}}\hskip.2em
      \subfloat[RankSRGAN: $19.07$/$0.564$ \newline $0.294$ / $0.132$]{\includegraphics[height=0.175\linewidth]{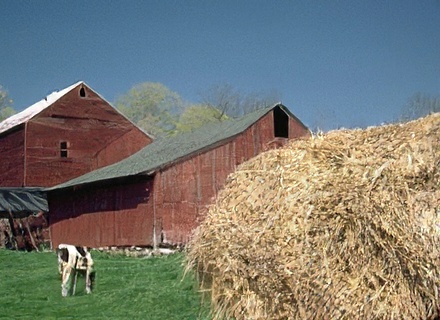}}
    \caption{A visual quality assessment example of super-resolution. (a) High-resolution image. (b)-(h) are the super-resolution results computed using bicubic interpolation, Glasner09~\citep{glasner2009super}, Yang13~\citep{yang2013fast}, EDSR~\citep{lim2017enhanced}, SRGAN~\citep{ledig2017photo}, ESRGAN~\citep{wang2018esrgan}, and RankSRGAN~\citep{zhang2019ranksrgan}, respectively. One can see that the GAN-based results (f)-(h) are visually superior to the others, contrary to the predictions of PSNR and SSIM. LPIPS indicates that the result (f) is worse than (d) and (e), in disagreement with visual inspection. DISTS is correlated well with human perception in this example.}
    \label{fig_ap:sr}
  \end{figure*}

  Fig.~\ref{fig_ap:sr} shows a quality assessment example of real-world super-resolution methods. Here we only compared the most widely used measures (PSNR and SSIM), and the two that performed best both on optimization and assessment (LPIPS and DISTS). It is not surprising that PSNR and SSIM have the poor correlation with human opinions, as they focus more on signal fidelity  than perceptual quality \citep{blau2018perception}. LPIPS and DISTS perform better, but the former is somewhat oversensitive to texture substitution.
  As many recent image restoration algorithms succeed in generating richer textures, DISTS  holds much promise for use in quality assessment for such applications.
  
  \begin{figure*}[t]
    \centering
     \includegraphics[height=0.39\linewidth]{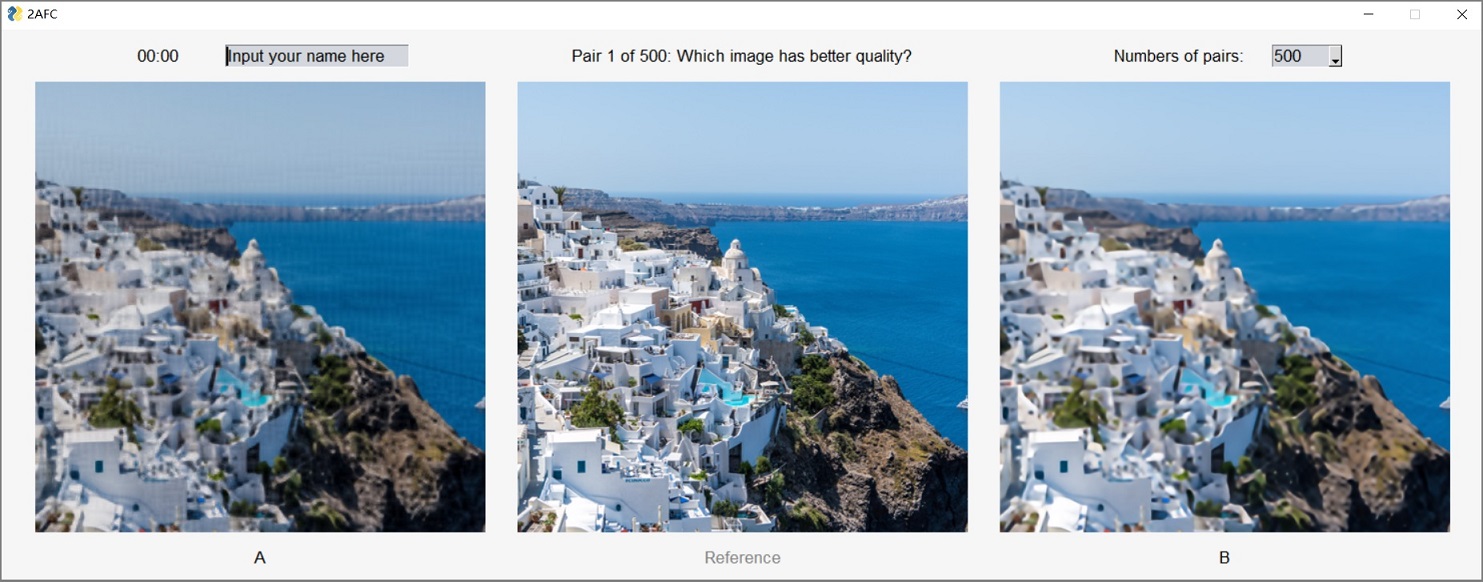}
    \caption{Customized graphical user interface for subjective testing.}
    \label{fig:gui}
  \end{figure*}

  \begin{figure*}[t]
    \centering
      \subfloat[Original: PSNR$\uparrow$ / SSIM$\uparrow$ \newline LPIPS$\downarrow$ / DISTS$\downarrow$] {\includegraphics[width=0.245\linewidth]{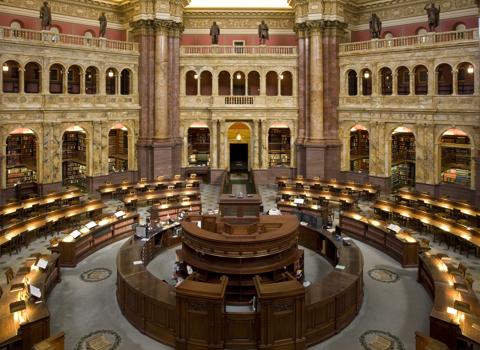}} \hskip.2em
      \subfloat[MAE: $26.44$ / $0.809$ \newline $0.285$ / $0.189$]{\includegraphics[width=0.245\linewidth]{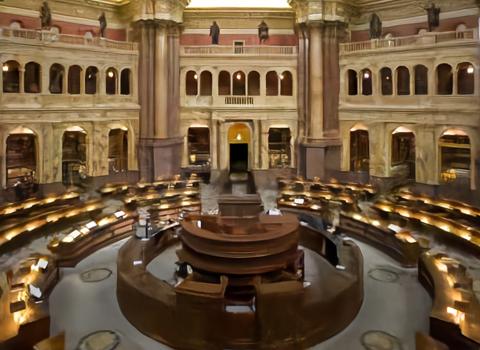}}\hskip.2em
      \subfloat[MS-SSIM: $26.28$ / $0.807$ \newline $0.286$ / $0.193$]{\includegraphics[width=0.245\linewidth]{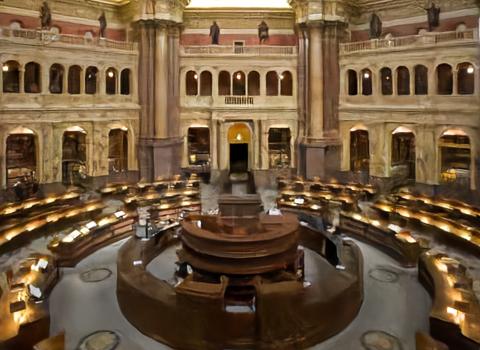}}\hskip.2em
      \subfloat[VIF: $26.18$ / $0.781$ \newline $0.303$ / $0.199$]{\includegraphics[width=0.245\linewidth]{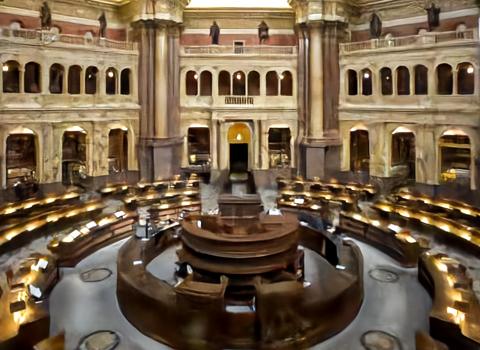}}\\ \vspace{-1em}
      \subfloat[CW-SSIM: $26.14$ / $0.787$ \newline $0.293$ / $0.184$]{\includegraphics[width=0.245\linewidth]{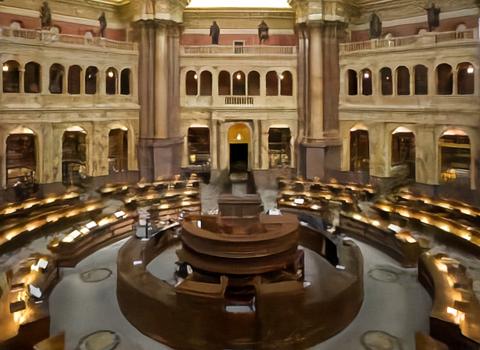}}\hskip.2em
      \subfloat[MAD: $26.11$ / $0.796$ \newline $0.288$ / $0.188$]{\includegraphics[width=0.245\linewidth]{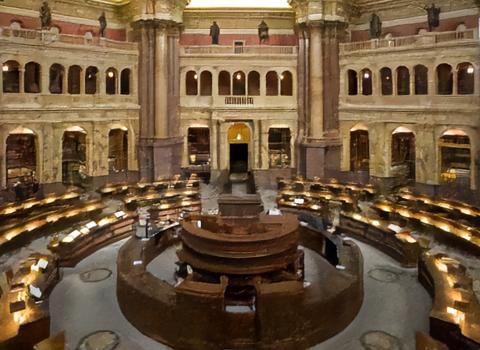}}\hskip.2em
      \subfloat[FSIM: $25.99$ / $0.784$ \newline $0.307$ / $0.209$]{\includegraphics[width=0.245\linewidth]{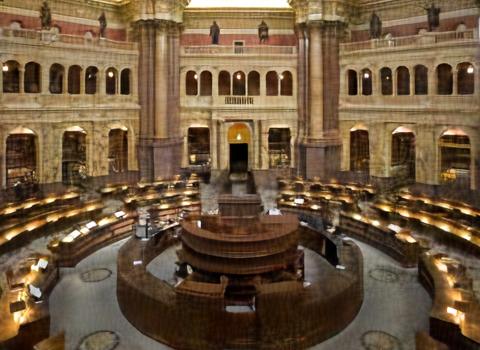}}\hskip.2em
      \subfloat[GMSD: $21.45$ / $0.707$ \newline $0.372$ / $0.285$]{\includegraphics[width=0.245\linewidth]{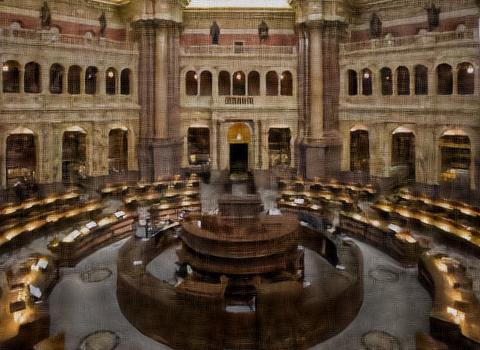}}\\ \vspace{-1em}
      \subfloat[VSI: $25.60$ / $0.784$ \newline $0.307$ / $0.216$]{\includegraphics[width=0.245\linewidth]{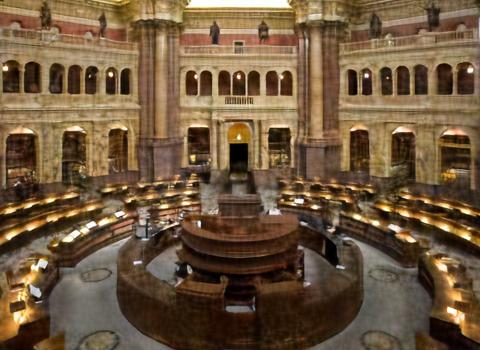}}\hskip.2em
      \subfloat[NLPD: $26.05$ / $0.795$ \newline $0.298$ / $0.201$]{\includegraphics[width=0.245\linewidth]{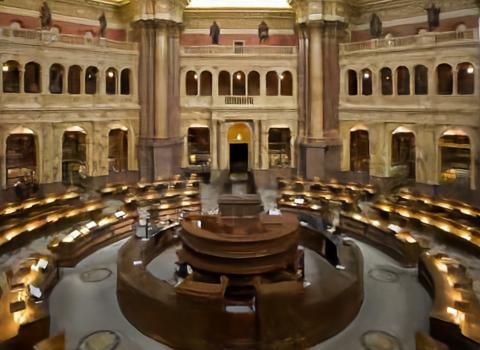}}\hskip.2em
      \subfloat[LPIPS: $25.55$ / $0.774$ \newline $0.285$ / $0.183$]{\includegraphics[width=0.245\linewidth]{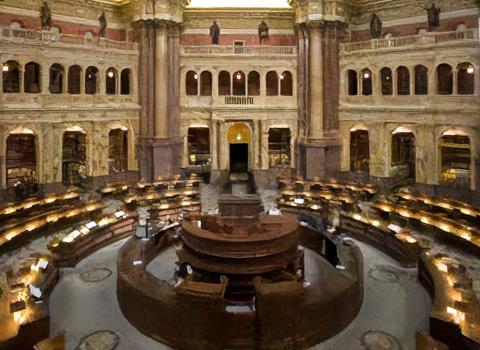}}\hskip.2em
      \subfloat[DISTS: $25.56$ / $0.775$ \newline $0.293$ / $0.175$]{\includegraphics[width=0.245\linewidth]{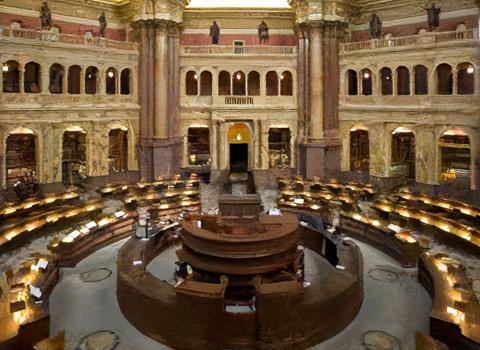}}
    \caption{Another set of denoising results optimized for different IQA models.}
    \label{fig:ap_n}
  \end{figure*}

  \begin{figure*}[t]
    \centering
      \subfloat[Original: PSNR$\uparrow$ / SSIM$\uparrow$ \newline LPIPS$\downarrow$ / DISTS$\downarrow$] {\includegraphics[width=0.245\linewidth]{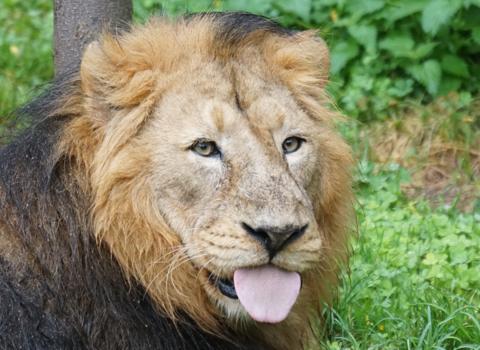}} \hskip.2em
      \subfloat[MAE: $22.91$ / $0.488$ \newline $0.458$ / $0.283$]{\includegraphics[width=0.245\linewidth]{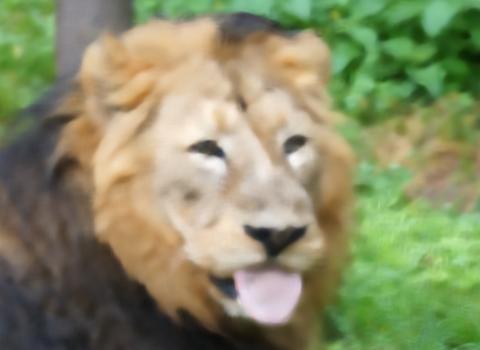}}\hskip.2em
      \subfloat[MS-SSIM: $22.68$ / $0.510$ \newline $0.430$ / $0.256$]{\includegraphics[width=0.245\linewidth]{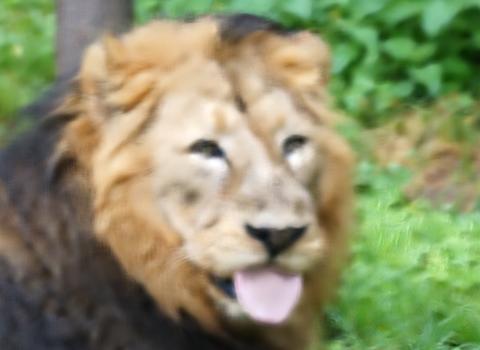}}\hskip.2em
      \subfloat[VIF: $15.96$ / $0.447$ \newline $0.441$ / $0.267$]{\includegraphics[width=0.245\linewidth]{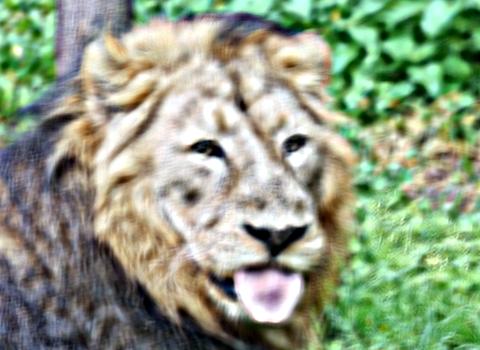}}\\ \vspace{-1em}
      \subfloat[CW-SSIM: $22.99$ / $0.493$ \newline $0.387$ / $0.192$]{\includegraphics[width=0.245\linewidth]{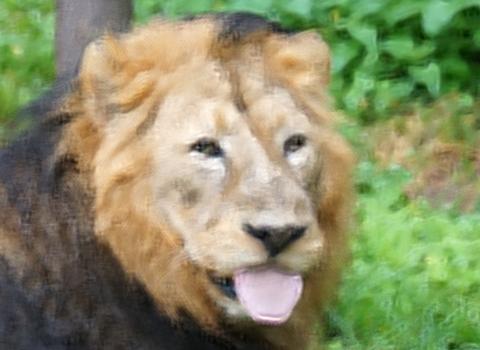}}\hskip.2em
      \subfloat[MAD: $22.07$ / $0.409$ \newline $0.463$ / $0.256$]{\includegraphics[width=0.245\linewidth]{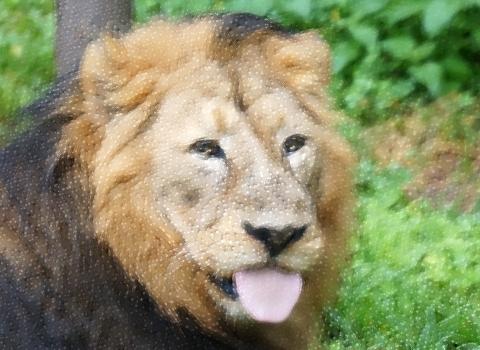}}\hskip.2em
      \subfloat[FSIM: $22.23$ / $0.382$ \newline $0.510$ / $0.323$]{\includegraphics[width=0.245\linewidth]{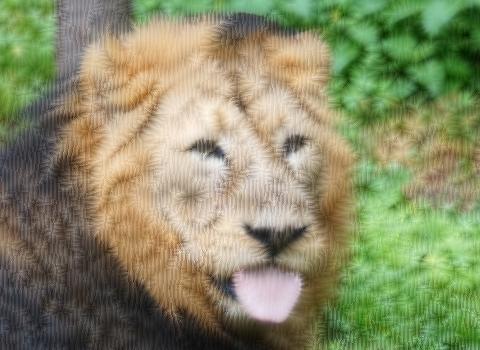}}\hskip.2em
      \subfloat[GMSD: $20.04$ / $0.401$ \newline $0.572$ / $0.371$]{\includegraphics[width=0.245\linewidth]{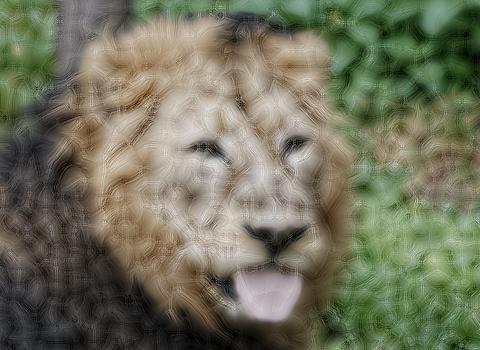}}\\ \vspace{-1em}
      \subfloat[VSI: $23.03$ / $0.435$ \newline $0.496$ / $0.381$]{\includegraphics[width=0.245\linewidth]{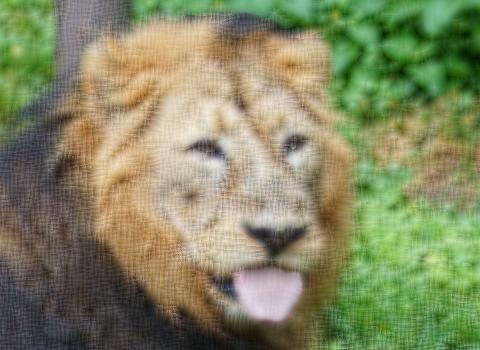}}\hskip.2em
      \subfloat[NLPD: $23.51$ / $0.477$ \newline $0.467$ / $0.294$]{\includegraphics[width=0.245\linewidth]{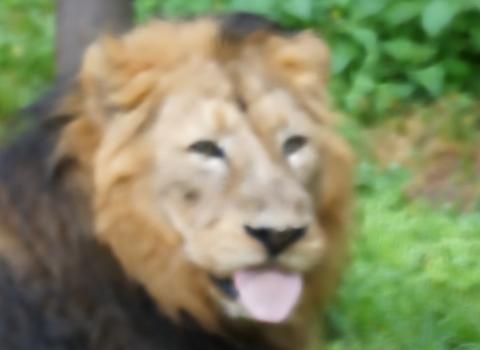}}\hskip.2em
      \subfloat[LPIPS: $22.50$ / $0.499$ \newline $0.272$ / $0.097$]{\includegraphics[width=0.245\linewidth]{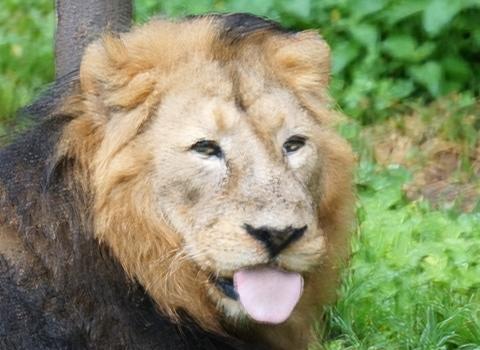}}\hskip.2em
      \subfloat[DISTS: $21.92$ / $0.471$ \newline $0.305$ / $0.101$]{\includegraphics[width=0.245\linewidth]{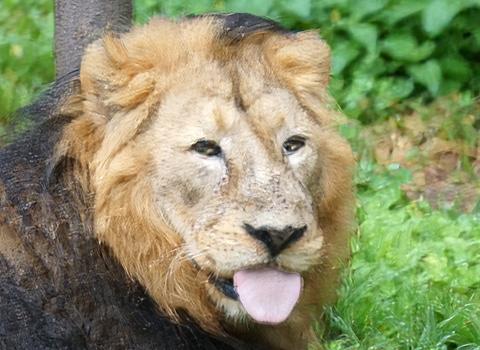}}
    \caption{Another set of deblurring results  optimized for different IQA models.}
    \label{fig:ap_b}
  \end{figure*}

  \begin{figure*}[t]
    \centering
      \subfloat[Original: PSNR$\uparrow$ / SSIM$\uparrow$ \newline LPIPS$\downarrow$ / DISTS$\downarrow$] {\includegraphics[width=0.245\linewidth]{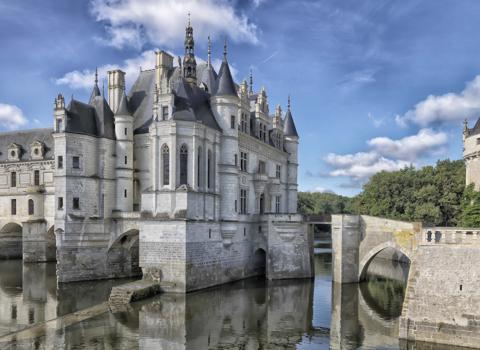}} \hskip.2em
      \subfloat[MAE: $27.16$ / $0.795$ \newline $0.328$ / $0.226$]{\includegraphics[width=0.245\linewidth]{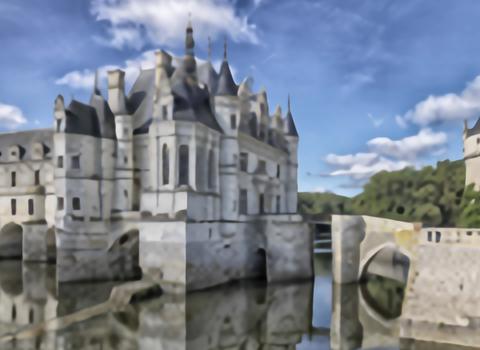}}\hskip.2em
      \subfloat[MS-SSIM: $27.01$ / $0.799$ \newline $0.321$ / $0.222$]{\includegraphics[width=0.245\linewidth]{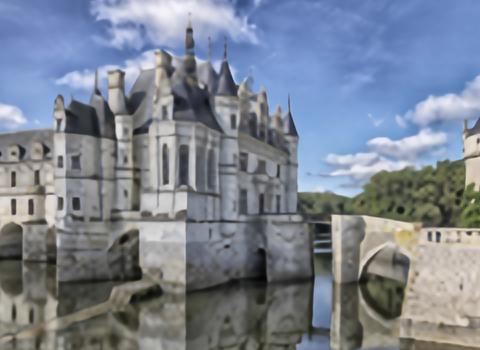}}\hskip.2em
      \subfloat[VIF: $17.72$ / $0.659$ \newline $0.432$ / $0.298$]{\includegraphics[width=0.245\linewidth]{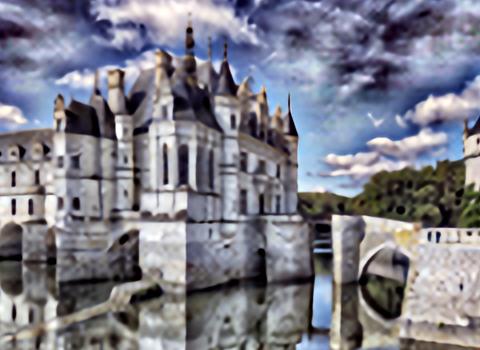}}\\ \vspace{-1em}
      \subfloat[CW-SSIM: $25.79$ / $0.738$ \newline $0.369$ / $0.258$]{\includegraphics[width=0.245\linewidth]{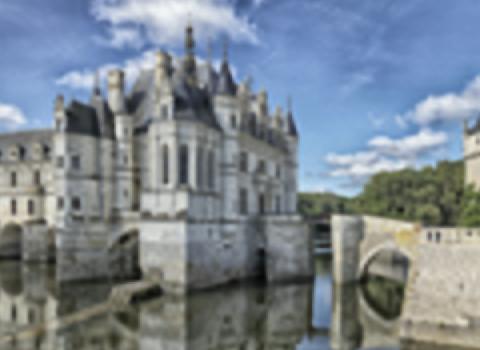}}\hskip.2em
      \subfloat[MAD: $25.96$ / $0.753$ \newline $0.313$ / $0.179$]{\includegraphics[width=0.245\linewidth]{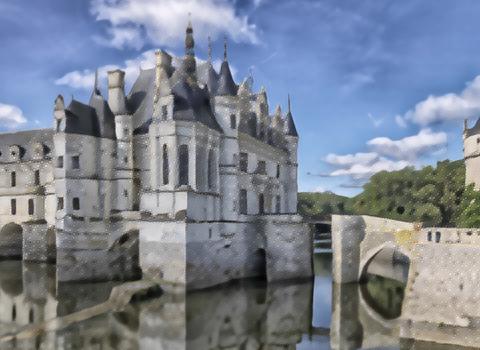}}\hskip.2em
      \subfloat[FSIM: $26.11$ / $0.766$ \newline $0.313$ / $0.197$]{\includegraphics[width=0.245\linewidth]{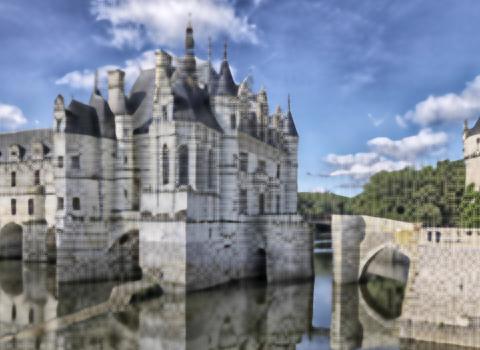}}\hskip.2em
      \subfloat[GMSD: $20.56$ / $0.750$ \newline $0.321$ / $0.215$]{\includegraphics[width=0.245\linewidth]{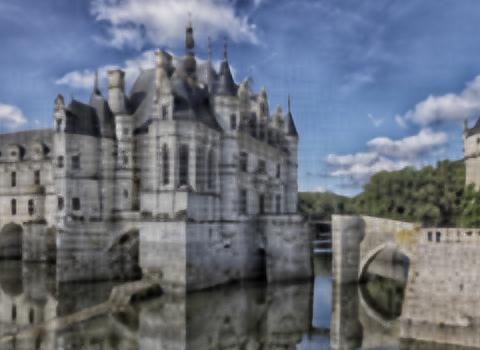}}\\ \vspace{-1em}
      \subfloat[VSI: $25.87$ / $0.755$ \newline $0.336$ / $0.232$]{\includegraphics[width=0.245\linewidth]{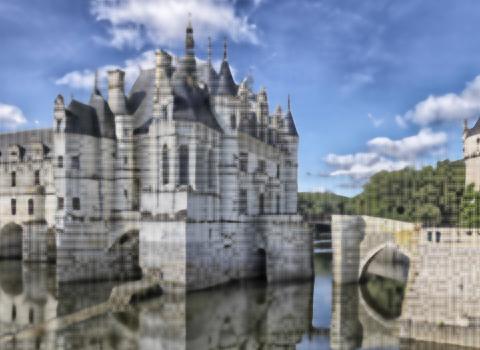}}\hskip.2em
      \subfloat[NLPD: $27.16$ / $0.793$ \newline $0.324$ / $0.224$]{\includegraphics[width=0.245\linewidth]{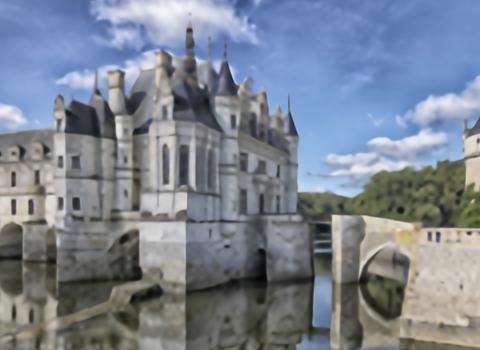}}\hskip.2em
      \subfloat[LPIPS: $25.90$ / $0.758$ \newline $0.219$ / $0.123$]{\includegraphics[width=0.245\linewidth]{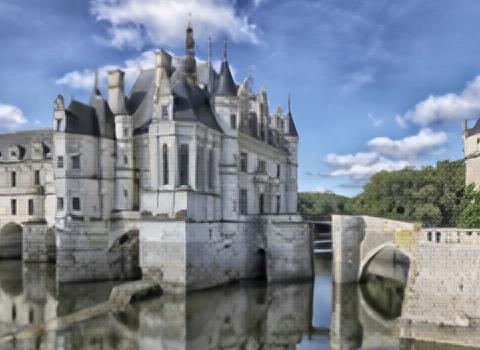}}\hskip.2em
      \subfloat[DISTS: $25.22$ / $0.740$ \newline $0.236$ / $0.107$]{\includegraphics[width=0.245\linewidth]{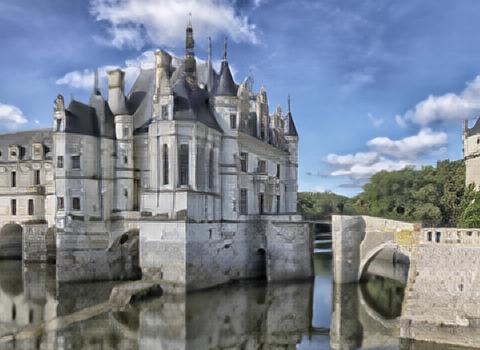}}
    \caption{Another set of  super-resolution results optimized for different IQA models.}
    \label{fig:ap_s}
  \end{figure*}
  
  \begin{figure*}[t]
    \centering
      \subfloat[Original: PSNR$\uparrow$ / SSIM$\uparrow$ \newline LPIPS$\downarrow$ / DISTS$\downarrow$] {\includegraphics[width=0.245\linewidth]{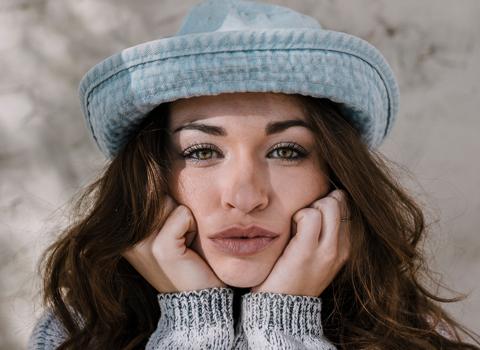}} \hskip.2em
      \subfloat[MAE: $27.55$ / $0.834$ \newline $0.351$ / $0.209$]{\includegraphics[width=0.245\linewidth]{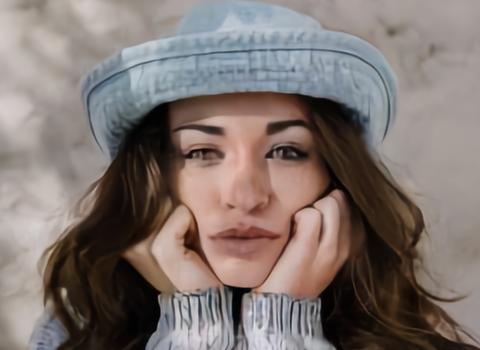}}\hskip.2em
      \subfloat[MS-SSIM: $27.67$ / $0.846$ \newline $0.320$ / $0.189$]{\includegraphics[width=0.245\linewidth]{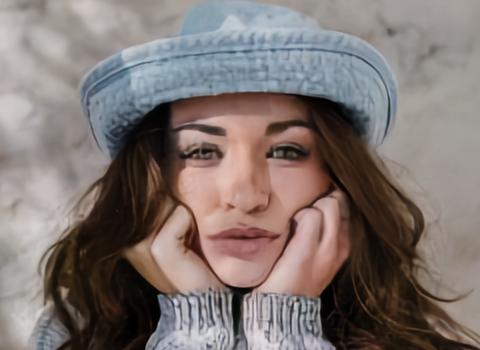}}\hskip.2em
      \subfloat[VIF: $14.75$ / $0.575$ \newline $0.449$ / $0.290$]{\includegraphics[width=0.245\linewidth]{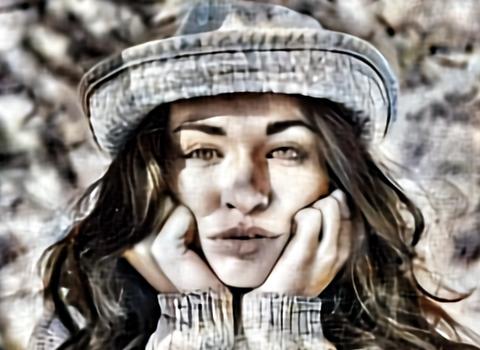}}\\ \vspace{-1em}
      \subfloat[CW-SSIM: $22.42$ / $0.429$ \newline $0.613$ / $0.407$]{\includegraphics[width=0.245\linewidth]{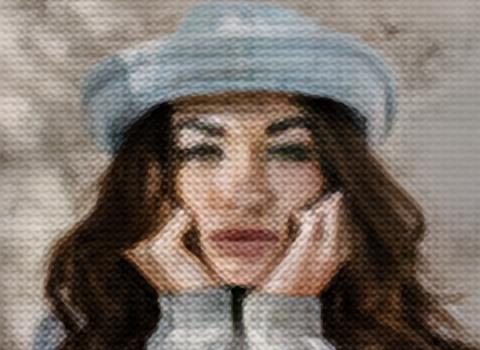}}\hskip.2em
      \subfloat[MAD: $26.55$ / $0.793$ \newline $0.331$ / $0.181$]{\includegraphics[width=0.245\linewidth]{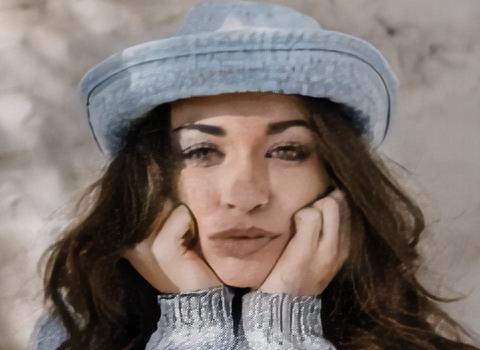}}\hskip.2em
      \subfloat[FSIM: $25.92$ / $0.831$ \newline $0.324$ / $0.184$]{\includegraphics[width=0.245\linewidth]{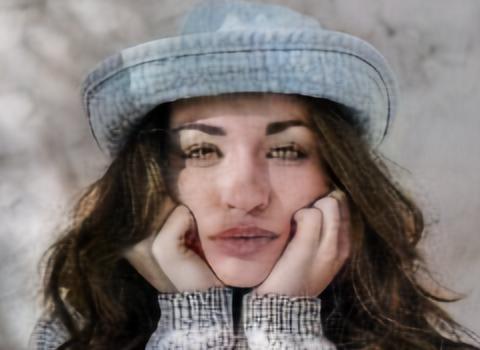}}\hskip.2em
      \subfloat[GMSD: $21.99$ / $0.762$ \newline $0.388$ / $0.197$]{\includegraphics[width=0.245\linewidth]{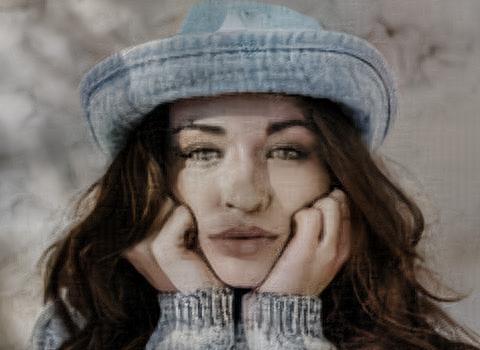}}\\ \vspace{-1em}
      \subfloat[VSI: $25.14$ / $0.794$ \newline $0.380$ / $0.220$]{\includegraphics[width=0.245\linewidth]{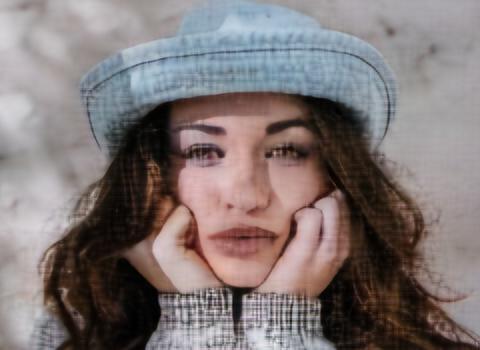}}\hskip.2em
      \subfloat[NLPD: $21.83$ / $0.762$ \newline $0.353$ / $0.201$]{\includegraphics[width=0.245\linewidth]{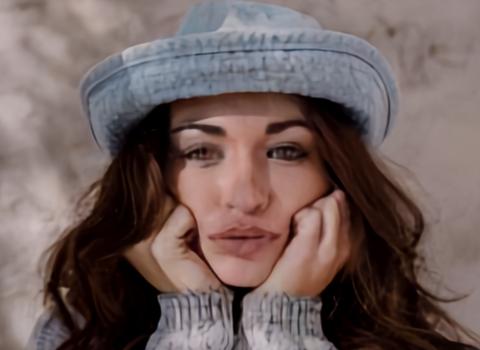}}\hskip.2em
      \subfloat[LPIPS: $24.36$ / $0.786$ \newline $0.161$ / $0.081$]{\includegraphics[width=0.245\linewidth]{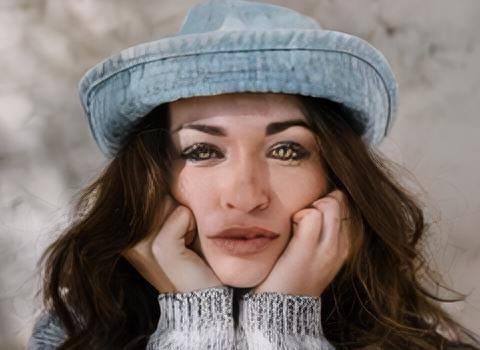}}\hskip.2em
      \subfloat[DISTS: $25.18$ / $0.781$ \newline $0.201$ / $0.080$]{\includegraphics[width=0.245\linewidth]{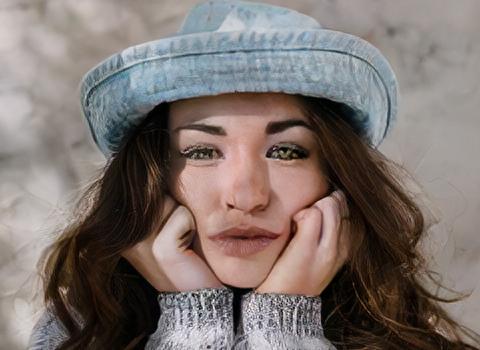}}
    \caption{Another set of  compression results optimized for different IQA models.}
    \label{fig:ap_c}
  \end{figure*}

  \end{appendices}

  

  %
  %

  \bibliographystyle{spbasic}      


  \end{document}